\documentclass{elsarticle}

\usepackage{hyperref} 

\journal{Mechanical Systems and Signal Processing}

\usepackage{amssymb}
\usepackage{graphics}
\usepackage{amsmath,bm}
\usepackage{subfigure}
\usepackage{epstopdf} 
\usepackage[bbgreekl]{mathbbol}
\usepackage{color}
\usepackage[compatibility=false]{caption}
\usepackage{placeins}
\usepackage{setspace}
\usepackage{xcolor}
\definecolor{test}{rgb}{0.7,0.7,1}
\definecolor{test2}{rgb}{1,0.7,0.7}
\usepackage{mathrsfs}
\usepackage{amsmath}
\usepackage{stmaryrd}
\usepackage{placeins}
\usepackage{mathtools}
\usepackage{placeins}
\usepackage{tikz}
\usepackage{etoolbox}

\newrobustcmd*{\mysquare}[1]{\tikz{\filldraw[draw=#1,fill=#1] (0,0) rectangle (0.2cm,0.2cm);}}

\newrobustcmd*{\mycircle}[1]{\tikz{\filldraw[draw=#1,fill=#1] (0,0) circle [radius=0.1cm];}}

\newrobustcmd*{\mytriangle}[1]{\tikz{\filldraw[draw=#1,fill=#1] (0,0) -- (0.2cm,0) -- (0.1cm,0.2cm);}}

\newcommand\BibTeX{{\rmfamily B\kern-.05em \textsc{i\kern-.025em b}\kern-.08em
		T\kern-.1667em\lower.7ex\hbox{E}\kern-.125emX}}

\renewcommand{\SS}{\Theta}

\newcommand{\SSpt}{\theta}

\newcommand{\SA}{\mathbb{\Sigma}}

\newcommand{\PM}{\mathbb{P}}

\newcommand{\randvar}[1]{\mathbb{#1}}

\newcommand{\pdf}[1]{p_{{\tiny #1}}}

\newcommand{\mean}[1]{\mu_{#1}}

\newcommand{\R}{\ensuremath{\mathbb{R}}}           
\newcommand{\Z}{\ensuremath{\mathbb{Z}}}           

\begin{document}

\begin{frontmatter}

\title{Damage detection in uncertain nonlinear systems based on stochastic Volterra series}

\author{Luis G. G. Villani}\ead{luis.villani@unesp.br}
\author{ Samuel da Silva}\ead{samuel.silva13@unesp.br}
\address{UNESP - Universidade Estadual Paulista, Faculdade de Engenharia de Ilha Solteira, Departamento de Engenharia Mec\^anica, Av. Brasil, 56, Ilha Solteira, 15385-000, SP, Brazil}
\author{Americo Cunha Jr.}\ead{americo@ime.uerj.br}
\address{UERJ - Universidade do Estado do Rio de Janeiro, NUMERICO -- Nucleus of Modeling and Experimentation with Computers, R. S\~ao Francisco Xavier, 524, Rio de Janeiro, 20550-900, RJ, Brazil}


\cortext[mycorrespondingauthor]{Luis Gustavo Giacon Villani}



\begin{abstract}
The damage detection problem in mechanical systems, using vibration measurements, is commonly called \textit{Structural Health Monitoring} (SHM). Many tools are able to detect damages by changes in the vibration pattern, mainly, when damages induce nonlinear behavior. However, a more difficult problem is to detect structural variation associated with damage, when the mechanical system has nonlinear behavior even in the reference condition. In these cases, more sophisticated methods are required to detect if the changes in the response are based on some structural variation or changes in the vibration regime, because both can generate nonlinearities. Among the many ways to solve this problem, the use of the Volterra series has several favorable points, because they are a generalization of the linear convolution, allowing the separation of linear and nonlinear contributions by input filtering through the Volterra kernels. On the other hand, the presence of uncertainties in mechanical systems, due to noise, geometric imperfections, manufacturing irregularities, environmental conditions, and others,  can also change the responses, becoming more difficult the damage detection procedure. An approach based on a stochastic version of Volterra series is proposed to be used in the detection of a breathing crack in a beam vibrating in a nonlinear regime of motion, even in reference condition (without crack). The system uncertainties are simulated by the variation imposed in the linear stiffness and damping coefficient. The results show, that the nonlinear analysis done, considering the high order Volterra kernels, allows the approach to detect the crack with a small propagation and probability confidence, even in the presence of uncertainties.
\end{abstract}

\begin{keyword}
\texttt{Damage detection \sep uncertainties quantification \sep 
	stochastic Volterra series \sep nonlinear dynamics.}
\end{keyword}

\end{frontmatter}


\section{Introduction}
\label{Introduction}
The process of damage detection in mechanical, aerospace and civil systems and structures, based on damage features and statistical analysis of them, is commonly called as \textit{Structural Health Monitoring} (SHM) \cite{Farrar2007}. The study and application of SHM techniques are motivated by the potential life-safety and economic impact of their implementation, which contributes to the development of new approaches \cite{farrar2012structural}. As it is widely known, the damage-sensitive features can be sensitive to confounding effects such as environmental or operating conditions, temperature and humidity changes, sensor bonding conditions, and others \cite{WORDEN2018139}. In addition, real systems and structures are subject to uncertainties, mostly related to noise in the measurements (experimental data variation), geometric imperfections, manufacturing irregularities, environmental conditions, or even, lack of knowledge about the system physics \cite{Soize20132379,soize2012stochastic,Soize2017}. These uncertainties can complicate the damage detection problem and have to be considered in the SHM applications, reducing the occurrence of false alarms \cite{MAO2013333}. 

To overcome this problem, the damage detection can be done based on statistical procedures, taking into account confidence limits to the dynamic behavior of the systems and structures. Thus, the structural monitoring can be done, usually, based on: (i) novelty detection, i. e., detection of  discrepant behaviors of the structure compared with the normal condition which it is known, in most cases with the establishment of statistical thresholds to the system behavior in normal condition; (ii) classification, in this case all states are known and the condition of the structure or system is classified into groups of different damage stages or healthy condition; (iii) regression algorithms, in which some variable or process is monitored with confidence limits. The choice of the more interesting approach to be used it depends on the specific problem and on the level of knowledge about the damage (detection, location, assessment, prediction) desired  \cite{WORDEN20151}. In this sense, a lot of methods can be used, such as linear and nonlinear Principal Component Analysis (PCA - NPCA) \cite{Pavlopoulou2016} , Extreme Value Statistics (EVS) \cite{sohn2005structural}, Peaks Over Threshold (POT) \cite{MarcRebillat2016}, machine learning algorithms \cite{SANTOS2016584}, neural network \cite{WORDEN2003323}, Bayesian approaches \cite{FIGUEIREDO20141}, Mahalanobis distance \cite{WORDEN2000647,FIGUEIREDO2010822}, and others. In all these applications, the final objective of the methods used are the same, differentiate the uncertainties and variabilities of the damages.    

On the other hand, many existing SHM features use the nonlinear characteristics of damages to detect them. In this context, cracks \cite{LIM2014468}, delamination \cite{Mandal2015} or rubbing in rotor systems \cite{ZENG2015380}, induce linear systems to exhibit nonlinear phenomena and the damage detection problem becomes a problem of nonlinearities detection. So, in this situation, the damage can be detected using harmonic distortions, coherence functions, probability density functions, correlation tests, Hilbert transform and others \cite{STC:STC215}.  Thus, the crack detection in beams can be performed based on the nonlinear behavior induced by this type of damage. Andreaeus et al. \cite{ANDREAUS2012382} have used the nonlinear aspects of the dynamic response of a beam under harmonic excitation to relate the nonlinear resonances to the presence and size of the crack. R\'ebillat et al. \cite{REBILLAT2014247} have used cascade of Hammerstein models to detect nonlinear damage through the monitoring of the energy of the nonlinear components in the system response. The approach was applied in a simulated breathing crack problem detection and in the monitoring of a composite plate.  Lim et al. \cite{LIM2014468} have shown a reference-free fatigue crack detection technique, based on nonlinear ultrasonic modulation. In this last work,  a sequential outlier analysis was used to identify the crack
presence without referring to any baseline data obtained from the intact condition of the structure.

All these approaches are powerful tools used to detect the presence of fatigue crack or any other damage that induces nonlinear effects in the system behavior. However, the SHM problem increases when the intrinsic nonlinear behavior of the systems and structures is considered because the nonlinear phenomena can be confused with damage when classical SHM techniques, based on linear metrics, are used \cite{Bornn2010909}. Furthermore, many mechanical systems and structures can operate with strong nonlinear behavior, that makes them exhibit complex responses containing subharmonic and superharmonic resonances, jumps,  modal interactions, bifurcation, quasi-periodicity, and possible chaos \cite{Virgin,noel2017nonlinear}. In this situation, the nonlinear behavior of the system in the reference condition has, necessarily, to be considered. 

In this regard, many approaches, in time and frequency domain, can be used to describe the nonlinear behavior, such as Hilbert transform, Narmax Models, High-Order FRFs \cite{kerschen2007nonlinear,worden2007nonlinear}, restoring force surface methods (RFS) \cite{masri1979nonparametric,Noel2012}, harmonic balance method \cite{Tang16092015} and others. Unfortunately, these nonlinearities have many individualities, in a way that it is difficult to obtain a general model that describes all the structures of interest.  
Among the different methods for nonlinearities identification, the Volterra series stands out, because it is a generalization of the linear convolution concept, allowing the separation of the system response in linear and nonlinear components \cite{Schetzen}. The main procedure to estimate the Volterra kernels is the Harmonic Probing method, that was extensively used in system identification problems \cite{CHATTERJEE20101716,Scussel2017} and damage characterization \cite{Chatterjee20103325}. The limitation of the approach is the dependence of a parametric model of the system. 

Another possible formulation is the direct use of input/output signals to estimate the Volterra kernels. Tang et al. (2010) \cite{TANG20101099} have used the classical version of the Volterra series approach to detect structural variations in a rotor, using the vibration signals of different points as input/output signals. Tang et al. (2010) have used the classical version of the Volterra series approach to detect structural variations in a rotor, using the vibration signals of different points as input/output signals. The Volterra kernels coefficients and the Principal Component Analysis (PCA) were used as damages features. In this study, the use of the first 3 kernels has shown better performance, with the capability of separate all the conditions considered. However, as mentioned by the authors, the high number of parameters to be identified has difficulted the approach implementation using an optimization process. Similarly,  Xia et al. (2016) \cite{XIA2016557} have used a key kernels-PSO (KK-PSO) method to identify the Volterra model, combined with neural networks, also to detect structural variation in a rotor. The implementation has been facilitated by the model reduction obtained using the KK-PSO method.  The results have shown the capability to the method to detect different types of structural variation. In both works, the Volterra series coefficients have been used as damage sensitivity feature, but they have not considered the presence of uncertainties in the damage detection process to study the capability of the approaches to detect the damages with a probabilistic confidence.

Additionally, Shiki et al. (2017) \cite{BruceSHMjournal} have used this Volterra series formulation, based on input/output signals, to detect structural variations related with damage, in a beam with nonlinear behavior, even in the reference condition. They have shown the use of Volterra series expanded through the Kautz functions, as a reference model to predict the nonlinear system response and to monitor the structure of interest. A statistical study of the damage index proposed was done with satisfactory results, but only the variation related to measurement noise was taken into account. The authors have not considered the possible system response variation related to the uncertainties, like environmental and operating conditions, and others as mentioned before.     

Thus, in this work, it is proposed to use a stochastic version of the Volterra series approach, expanded using random Kautz functions, as a stochastic model for damages detection based on novelty detection. The Volterra kernels are estimated in a probabilistic framework, considering the data variation, that is obtained through the variabilities in fundamental frequency and damping of the studied system. To allow the description of the response variation, the Kautz parameters are treated as random variables and the Kautz functions, used to reduce the number of terms of the Volterra kernels, as random processes. This methodology gives to the proposed model the ability to describe the system nonlinear behavior and the data variation simultaneous. 

To detect the presence of damage, the stochastic model obtained through the stochastic version of Volterra series is used as a set of reference models, so new models are estimated in unknown condition and compared with this set of models, through a distance-based method. The capability of the Volterra series model to separate linear and nonlinear contributions in system response is used. Two methodologies are compared, considering the linear and nonlinear components of the Volterra series. The first methodology considers only the Volterra kernels coefficients and the second considers the kernels contribution. The approaches are applied in a simulated uncertain nonlinear system that describes the real behavior of a nonlinear beam, and, the damage is simulated by the propagation of a breathing crack, described as a bilinear stiffness oscillator. The results obtained show that the methodologies proposed are able to differentiate the nonlinear behavior and data variation of the damage presence, with better performance than the simple linear analysis. It is showed that the use of Volterra series in this situation represents a clear advantage in damage detection process, with the improvement in the results in comparison with the linear approach. In addition, the stochastic version proposed allows the damage detection through the approach, even in the presence of uncertainties, with probability confidence. 

The present paper is organized as follows. Section \ref{The Sotochastic Volterra Series Approach for Damage Detection} reviews the fundamental aspects of system identification based on Volterra series, showing the expansion of the theory to the proposed stochastic Volterra series model, and describes the methodology used to detect damage in nonlinear systems, considering uncertainties of different natures. Section \ref{Nonlinear System of Interesting} presents the model of the nonlinear uncertain beam simulated and the breathing crack model used to simulate the presence and propagation of damage. Section \ref{Numerical Results} shows the results obtained through the application of the approach in the studied system. Finally, section \ref{Final Remarks} presents the main conclusions and some paths for future work.

\section{The stochastic Volterra series approach for damage detection}
\label{The Sotochastic Volterra Series Approach for Damage Detection}
This section describes the proposed methodology for modeling nonlinear systems assuming data uncertainties. The approach is based on a stochastic version of the Volterra series, expanded using Kautz functions as the orthonormal basis. The sections \ref{determ_volt_series} and \ref{determ_kautz} summarize the deterministic version of the Volterra series expanded using Kautz functions,  which is known in the literature and it was used to detect damage in initially nonlinear systems by Shiki et al. \cite{BruceSHMjournal}. Additionally, the sections \ref{System uncertainties}, \ref{The stochastic Volterra series} and \ref{The stochastic Kautz parameters} show the expansion of the classical deterministic concept  to the new random version of the approach, proposed to be used to model uncertain nonlinear systems. Finally, section \ref{Damage detection using Volterra series} presents the two new proposed damage detection methodologies, to be used considering the random version of the Volterra series.

\subsection{Deterministic Volterra series}
\label{determ_volt_series}
Consider a discrete-time causal nonlinear system with a single output
$k \in \Z_+ \mapsto y(k)$ caused by a single input 
$k \in \Z_+ \mapsto u(k)$, with $\Z_+$ representing the set of nonnegative integers. Through the discrete-time Volterra series, the output of this nonlinear system can be written in the form
\begin{equation}
y(k) = \sum_{\eta = 1}^\infty\sum_{n_{1}=0}^{N_{1}-1} \ldots \sum_{n_{\eta}=0}^{N_{\eta}-1} 
\mathcal{H}_{\eta}(n_{1},\ldots,n_{\eta}) \prod_{i=1}^{\eta}u(k-n_{i}),
\label{volt_series_eq}
\end{equation}
where $(n_1,..,n_\eta) \in \Z^\eta _+ \mapsto \mathcal{H}_{\eta}(n_{1},\ldots,n_{\eta})$ 
represents the $\eta$-order Volterra kernel.
This formalism is very versatile, in a way that it allows one to represent 
different types of nonlinear systems using the convolution concept 
\cite{Schetzen}. Also, it allows one to split the system output 
into a sum of linear and nonlinear contributions
\begin{equation}
y(k) = \underbrace{y_1(k)}_{linear} + \underbrace{y_2(k) + y_3(k) + \cdots + y_{\eta}(k)}_{nonlinear}.
\end{equation}
The main drawback of the approach is the difficulty of convergence of the series when a large number of terms $N_1,...,N_\eta$ are used.
In order to reduce the number of terms necessary to obtain a good approximation, 
the Volterra kernels $\mathcal{H}_{\eta}$ can be expanded into an orthonormal basis, such as 
the Kautz functions, that is employed in this work \cite{Kautz,heuberger2005modelling}. 
In this way, one can write
\begin{eqnarray}
\mathcal{H}_{\eta}(n_{1},...\, ,n_{\eta}) \approx \sum_{i_{1}=1}^{J_{1}}...\sum_{i_{\eta}=1}^{J_{\eta}}
\mathcal{B}_{\eta} \left(i_{1},...\, ,i_{\eta} \right) \prod_{j=1}^{\eta} \psi_{\eta, i_{j}} (n_{j})\,,
\label{kautz_aprox_eq}
\end{eqnarray}
\noindent where $J_{1},\ldots,J_{\eta}$ are the number of Kautz functions used in each orthonormal 
projections of the Volterra kernels, $(i_{1},\ldots,i_{\eta}) \in \Z^\eta _+ \mapsto \mathcal{B}_{\eta}(i_{1},\ldots,i_{\eta})$ 
represents the $\eta$-order Volterra kernel, expanded in the orthonormal basis,
and $n_{j} \in \Z_{+} \mapsto \psi_{\eta, i_{j}} (n_{j})$ represents the $i_{j}$-th Kautz filter. 

Thus, using the discrete Volterra series representation of Eq. (\ref{volt_series_eq})
and the Kautz  approximation (\ref{kautz_aprox_eq}), the system response
can be approximated by the multidimensional convolution between the
orthonormal kernels ${\mathcal{B}}_{\eta}$ and the input signal filtered by the
Kautz functions, i.e.,
\begin{equation}
y(k)\approx \sum_{\eta = 1}^\infty \sum_{i_{1}=1}^{J_{1}}\ldots\sum_{i_{\eta}=1}^{J_{\eta}}
\mathcal{B}_\eta\left(i_{1},\ldots,i_{\eta}\right) \prod_{j=1}^{\eta} l_{\eta, i_{j}}(k)\, ,
\label{eqresp}
\end{equation}
where $k \in \Z_{+} \mapsto l_{\eta, i_{j}} (k)$ is a simple filtering of the input signal $u(k)$ 
by the Kautz function $\psi_{\eta, i_{j}}$, i.e.,
\begin{equation}
l_{\eta, i_{j}}(k) = \sum_{n_{i}=0}^{V-1} \psi_{\eta, i_{j}}(n_{i})u(k-n_{i})\,,
\end{equation}
\noindent where $V=\max\{J_{1},\ldots,J_{\eta}\}$.

More information on the identification approach based on Volterra series 
can be found in \cite{daSilva20111103,daSilva2011312}. Details about 
Kautz functions are given in section~\ref{determ_kautz}.
The reader is also encouraged to see 
\cite{WAHLBERG1996693,Oliveira2012,BruceSHMjournal}.

\subsection{Deterministic Kautz functions}
\label{determ_kautz}
The generalized form of the Kautz functions is written as \cite{Wahlberg1994}
\begin{equation}
\Psi_{\eta,2j-1}(z)=\frac{z \, \sqrt{(1-b_\eta^{2})(1-c_\eta^{2})}}{z^{2}+b_\eta(c_\eta-1)z-c_\eta} \, \left[\frac{-c_\eta z^{2}+b_\eta(c_\eta-1)z+1}{z^2+b_\eta(c_\eta-1)z-c_\eta}\right]^{j-1}\,,
\label{Kautz2capkautz}
\end{equation}
and
\begin{equation}
\Psi_{\eta,2j}(z)=\Psi_{\eta,2j-1}(z)\frac{z-b_\eta}{\sqrt{1-b_\eta^{2}}}\, ,
\label{Kautz1capkautz}
\end{equation}
being the values of $b_\eta$ and $c_\eta$, respectively, defined by
\begin{equation}
b_\eta = \frac{(\mathcal{Z}_{\eta}+\bar{\mathcal{Z}_{\eta}})}{1+\mathcal{Z}_{\eta}\bar{\mathcal{Z}_{\eta}}}   
\label{b}
\end{equation}
and
\begin{equation}
c_\eta = -\mathcal{Z}_{\eta}\bar{\mathcal{Z}_{\eta}}\, ,
\label{c}
\end{equation}
where $\mathcal{Z}$ and  $\bar{\mathcal{Z}}$ are, respectively, the Kautz poles and its complex conjugate in discrete domain. The discrete poles $\mathcal{Z}_\eta$ can be related to the continuous poles $\mathcal{S}_\eta$ through the equation
\begin{equation}
\mathcal{Z}_\eta = \mbox{exp}\left(\frac{\mathcal{S}_\eta}{F_s}\right) \, ,
\end{equation}
where $F_s$ represents the sampling frequency. The Kautz poles for each kernel are defined as 
\begin{equation}
\mathcal{S}_{\eta}=-\xi_{\eta}\omega_{\eta} \pm j\omega_{\eta}\sqrt{1-\xi_{\eta}^{2}}\, ,
\label{pole1}
\end{equation}
\noindent
where $\omega_{\eta}$ and $\xi_{\eta}$ are the Kautz parameters and $\eta$ represents the number of the kernel considered. Remembering that, for the linear kernel, the Kautz parameters are the natural frequency and damping ratio of the system. In identification processes, it is common to use some optimization methodology to find $\omega_{\eta}$ and $\xi_{\eta}$ \cite{AdaRosa}. The use of the Kautz functions as the orthonormal basis is driven by their properties to describe the oscillatory dynamic systems \cite{Kautz}.

\subsection{System uncertainties} 
\label{System uncertainties}
Uncertainties are classified in the literature as being of two types, aleatory or epistemic. The first type is intrinsic to scenarios with variabilities, such as measurement noise in the system observations (experimental data) and variabilities of the real system with respect to its nominal configuration (due to geometric imperfections, manufacturing irregularities, environmental conditions, etc.) \cite{Soize20132379,soize2012stochastic,Soize2017}. 
These uncertainties can not be eliminated, only better characterized. On the other hand, epistemic uncertainties are due to the lack of knowledge (ignorance) about the system physics. By increasing knowledge about a certain system, these uncertainties can be mitigated \cite{Soize20132379,soize2012stochastic,Soize2017}. 

For the sake of simplicity, in this work, only the aleatory uncertainties, also known as data uncertainties, are taken into account. In the context of system identification, such uncertainties are materialized in the form of variations in the model parameters. Implicit in this approach is the hypothesis that the Volterra series is capable of producing a reliable representation of the system response. 

\subsection{The stochastic Volterra series}
\label{The stochastic Volterra series}
In this work, a stochastic version of the Volterra series is proposed  to identify the nonlinear system of interest. The model parameters subjected to uncertainties are described as random variables or processes, defined on the probability space $(\SS, \SA, \PM)$, where $\SS$ is a sample space, $\SA$ is a $\sigma$-algebra over $\SS$, and $\PM$ is a probability measure. It is assumed that any random variable $\SSpt \in \SS \mapsto \randvar{Y} (\SSpt) \in \R$ in this probabilistic setting, with a probability distribution $P_{\randvar{Y}}(dy)$ on $\R$, admits a probability density function (PDF) $y \mapsto p_{\randvar{Y}}(y)$ with respect to $dy$. 

Therefore, the discrete-time Volterra series describes the random nonlinear system output as
\begin{equation}
\randvar{y}(\SSpt,k)=\sum_{\eta = 1}^\infty\sum_{n_{1}=0}^{N_{1}-1}\ldots\sum_{n_{\eta}=0}^{N_{\eta}-1}\randvar{H}_{\eta}(\SSpt,n_{1},\ldots,n_{\eta}) \prod_{i=1}^{\eta}u(k-n_{i})\, ,
\label{volt_series_eq_stochastic}
\end{equation}
\noindent where $u(k)$ is the deterministic input signal, the random process $(\SSpt,k) \in \SS \times \Z_{+} \mapsto \randvar{y}(\SSpt,k)$ is the aleatory system response and $(\SSpt,n_1,..,n_\eta) \in \SS \times \Z^\eta \mapsto \randvar{H}_{\eta}(\SSpt,n_{1},\ldots,n_{\eta})$ represents the random version of the $\eta$-order Volterra kernel. 

The kernels expansion in terms of Kautz functions now is written as
\begin{eqnarray}
\randvar{H}_{\eta}(\SSpt,n_{1},...\, ,n_{\eta})\approx\sum_{i_{1}=1}^{J_{1}}...\sum_{i_{\eta}=1}^{J_{\eta}}\randvar{B}_\eta\left(\SSpt,i_{1},...\, ,i_{\eta}\right)\prod_{j=1}^{\eta}\bbpsi_{\eta, i_{j}}(\SSpt,n_{j})\,,
\label{kautz_aprox_eq_stochastic}
\end{eqnarray}
where $J_{1},\ldots,J_{\eta}$ are the number of Kautz functions used in each orthonormal projection of the Volterra kernels, the random process
$(\SSpt,i_{1},\ldots,i_{\eta}) \in \SS \times \Z_{+}^\eta \mapsto \randvar{B}_{\eta}(\SSpt,i_{1},\ldots,i_{\eta})$ 
represents the $\eta$-order random Volterra kernel, expanded in the orthonormal basis, and $(\SSpt,n_{j}) \in \SS \times \Z_{+} \mapsto \bbpsi_{\eta, i_{j}}(\SSpt,n_{j})$ represents 
the random version of the $i_{j}$-th Kautz filter related with the $\eta$-order random Volterra kernel. The Kautz functions are modeled as random processes, because their definition is associated with the dynamic system response $\randvar{y}(\SSpt,k)$ and depends on the damping ratio and natural frequency, which are subjected to uncertainties.

The stochastic version equivalent of the approximation shown in (\ref{eqresp}) is written
\begin{equation}
\randvar{y}(\SSpt,k)\approx \sum_{\eta = 1}^\infty \sum_{i_{1}=1}^{J_{1}}\ldots\sum_{i_{\eta}=1}^{J_{\eta}}\mathbb B_\eta\left(\SSpt,i_{1},\ldots,i_{\eta}\right)\prod_{j=1}^{\eta} \randvar{l}_{\eta, i_{j}}(\SSpt, k)\, ,
\end{equation}
where the random process $(\SSpt,k) \in \SS \times \Z_{+} \mapsto \randvar{l}_{\eta, i_{j}}(\SSpt, k)$ 
is a simple filtering of the deterministic input signal $u(k)$ by the random Kautz function $\bbpsi_{\eta, i_{j}}$, i. e.,
\begin{equation}
\randvar{l}_{\eta, i_{j}}(\SSpt, k) = \sum_{n_{i}=0}^{V-1}\bbpsi_{\eta, i_{j}}(\SSpt, n_{i})u(k-n_{i})\, ,
\end{equation}
for $V=\max\{J_{1},\ldots,J_{\eta}\}$.

\begin{figure}[!htb]
	\begin{center}
		\includegraphics[scale=0.7]{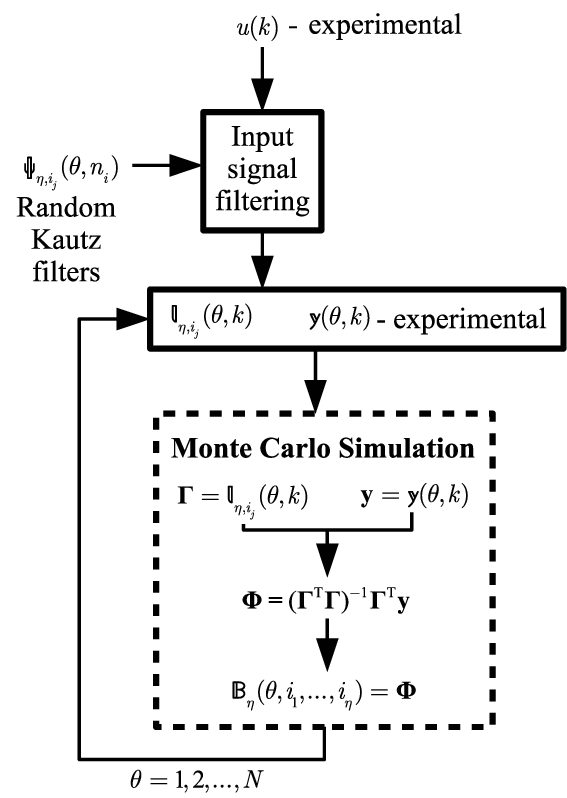}
		\caption{Description of the Volterra kernels identification approach used, based on Monte Carlo simulations.}
		\label{fluxoidenti}
	\end{center}
\end{figure}

Then, the coefficients of the kernels can be estimated, considering Monte Carlo simulations and the least squares method. Figure \ref{fluxoidenti} shows a flowchart of the Volterra kernels identification considering the random Kautz functions and the Monte Carlo simulations. So, considering each stochastic realization $\SSpt$, the matrix $\bm{{\Gamma}}$ can be completed with the input signal filtered $\randvar{l}_{\eta, i_{j}}(\SSpt, k)$ and the vector $\mathbf{{y}}$  with the experimental output signal $\randvar{y}(\SSpt,k)$
\begin{equation}
{\bm{{\Phi}}}=(\bm{{\Gamma}}^{T}\bm{{\Gamma}})^{-1}\bm{{\Gamma}}^{T}\mathbf{{y}}\, ,
\label{KernelsEstimation}
\end{equation}
\noindent where ${\bm{\Phi}}$ has the terms of the orthonormal kernels $\mathbb B_\eta$, in each realization $\SSpt$. The procedure is repeated until the Monte Carlo convergence is achieved. 

\subsection{The stochastic Kautz parameters} 
\label{The stochastic Kautz parameters}
In order to create a robust identification, it is necessary to provide some type of certification, i. e., an envelope of reliability for the nominal values of the model parameters. Such certification can be obtained by a stochastic model, where probability distributions, instead of scalar values, are identified for the model parameters. In this case, the system response also becomes random, a stochastic process to be precise, allowing the model to predict the variations in the response related to uncertainties. As consequence of the system response variation, the natural frequency and the damping ratio of the equivalent linear system can also vary, becoming random variables $\SSpt \in \SS \mapsto \bbzeta_n (\SSpt) \in \R$, $\SSpt \in \SS \mapsto \bbomega_n (\SSpt) \in \R$. As mentioned before, the Kautz parameters for each kernel are related with the systems natural frequency and damping ratio, so they are also considered as random variables $\SSpt \in \SS \mapsto \bbxi_\eta (\SSpt) \in \R$, $\SSpt \in \SS \mapsto \bbomega_\eta (\SSpt) \in \R$. 

So, the random system response is used to define the Kautz poles values. Thus, the Kautz parameters, related with the first Volterra kernel, can be approximated as natural frequencies and damping ratios of the equivalent linear system. These values can be obtained through modal analysis techniques and using the low amplitude system response, which is approximately linear. As the system linear dynamics vary, it is clear that the modal parameters, and consequently, the Kautz parameters vary too. The definition of the Kautz parameters related with the high order Volterra kernels is more complicated since they are not exactly natural frequencies and damping ratios. But, it is possible to define linear approximations between modal parameters and the high order Kautz parameters
\begin{eqnarray}
\bbomega_{1}(\SSpt) = \bbomega_{n}(\SSpt) \, , &\qquad&  \bbxi_{1}(\SSpt) = \bbzeta_n(\SSpt) \, , \nonumber\\
\bbomega_{2}(\SSpt) = p_1 \, \bbomega_{n}(\SSpt) \, , &\qquad& \bbxi_{2}(\SSpt) = p_2 \, \bbzeta_n(\SSpt) \, ,\label{Kautzrelationship}  \\ \bbomega_{3}(\SSpt) = p_3 \, \bbomega_{n}(\SSpt) \, , &\qquad& \bbxi_{3}(\SSpt) = p_4 \, \bbzeta_n(\SSpt) \, ,\nonumber
\end{eqnarray}
\noindent
where the relationships $p_1, p_2, p_3$ and $p_4$ are estimated minimizing the error function
\begin{eqnarray}
\underset{\Delta}{\mbox{min}} \, \, \textbf{J} = \sum_{k = 1}^{N} [\hat{y}(k) - y(k)]^2
\end{eqnarray}
\noindent where $y(k)$ is the deterministic discrete-time system response, $\hat{y}(k)$ is the deterministic model response and the vector $\Delta$ contains the coefficients of the relationships between modal and Kautz parameters. Once the coefficients were defined, the Kautz parameters variation can be obtained based on the model parameters variation. The optimization procedure does not need to be repeated every time, but only for one representative data. In the simulations performed, in this work, only the response correspondent to the deterministic values of the system parameters was used to optimize the Kautz poles. And, the variation of the Kautz poles was considered through the estimation of the modal parameters in each Monte Carlo realization.

\subsection{Damage detection based on Volterra series}
\label{Damage detection using Volterra series}
The stochastic Volterra series can be used as a mathematical model to approximate the system response, first in the healthy condition to define a set of reference models and, then, for each unknown condition (healthy or damaged), to make predictions about the condition of the system. Since the stochastic version of the Volterra series is able to describe linear and nonlinear systems, the damage detection approach can be applied with linear or nonlinear behavior in reference or damaged conditions. So, the system can be identified several times, using Monte Carlo simulations, in the reference condition, to construct a set of Volterra models. In the unknown condition, a new Volterra model can be identified and compared with the set of models constructed to detect the possible presence of structure variations related with damages, all this, considering the presence of uncertainties. The stochastic version of the Volterra series proposed allows the damage detection with probability confidence. In this work, two different approaches are shown, based on the kernels coefficients and on the linear and nonlinear contributions to the total system response, in each structural condition. The approaches are described as follow.

\subsubsection{Damage detection based on kernels coefficients}
\label{Damage detection based on Kernels coefficients}
The first approach is based on the coefficients of the Volterra kernels identified. This methodology does not consider the direct influence of the Kautz functions, because only the coefficients expanded in the orthonormal basis are considered. The influence of the random Kautz functions is taken into account only in the process of model identification. Considering Eq.~(\ref{KernelsEstimation}) and the Volterra series truncated in the third order term, the reduced order kernels coefficients, for each  Monte Carlo realization, are allocated as

\begin{footnotesize}
\begin{eqnarray}
\bm{\Phi} = \begin{bmatrix} \bm{\Phi_1} \\ \bm{\Phi_2} \\ \bm{\Phi_3} \end{bmatrix}, \quad
\bm{\Phi_1} = \begin{Bmatrix} \mathcal{B}_1(1) \\ \mathcal{B}_1(2) \\ \vdots \\ \mathcal{B}_1(J_1) \end{Bmatrix}, \quad
\bm{\Phi_2}  = \begin{Bmatrix} \mathcal{B}_2(1,1) \\ \mathcal{B}_2(1,2) \\ \vdots \\ \mathcal{B}_2(1,J_2) \\ \mathcal{B}_2(J_2,1) \\ \vdots \\ \mathcal{B}_2(J_2,J_2)\end{Bmatrix}, \quad
\bm{\Phi_3} = \begin{Bmatrix} \mathcal{B}_3(1,1,1) \\ \mathcal{B}_3(1,2,1) \\ \vdots \\ \mathcal{B}_3(1,J_3,1) \\ \mathcal{B}_3(1,1,2) \\ \vdots \\ \mathcal{B}_3(1,J_3,J_3)  \\ 
\vdots  \\ \mathcal{B}_3(J_3,J_3,J_3)\end{Bmatrix}\, ,
\end{eqnarray}
\end{footnotesize}

\noindent remembering that $J_1$, $J_2$ and $J_3$ are the number of Kautz functions used, considering the first, second and third kernels, respectively. It is expected that the main information about the high order kernels is allocated in the main diagonal, so it is considered as monitoring parameters only the coefficients positioned in this region. Then, the coefficients with higher contribution can be used to represent the model identified

\begin{footnotesize}
\begin{eqnarray}
\bm{\lambda}_{1} = \begin{Bmatrix} \mathcal{B}_1(1) \\  \mathcal{B}_1(2) \\ \vdots \\ \mathcal{B}_1(J_1) \end{Bmatrix}, \quad
\bm{\lambda}_{2} = \begin{Bmatrix} \mathcal{B}_2(1,1) \\ \mathcal{B}_2(2,2) \\ \vdots \\ \mathcal{B}_2(J_2,J_2) \end{Bmatrix}, \quad
\bm{\lambda}_{3} = \begin{Bmatrix} \mathcal{B}_3(1,1,1) \\ \mathcal{B}_3(2,2,2) \\ \vdots \\ \mathcal{B}_3(J_3,J_3,J_3) \end{Bmatrix}, \quad
\bm{\lambda}_{nl} = \begin{Bmatrix} \bm{\lambda}_{2} \\ \bm{\lambda}_{3} \end{Bmatrix}.
\end{eqnarray}
\end{footnotesize}

Now, considering the stochastic version of the Volterra series and the Monte Carlo simulations, multiple models are estimated in the reference condition and each vector becomes a random process. The random processes estimated, using Monte Carlo simulations, represents the set of models identified in reference condition

\begin{footnotesize}
\begin{eqnarray}
{\bblambda_{1}} = \begin{Bmatrix} \randvar{B}_1(\SSpt,1) \\  \randvar{B}_1(\SSpt,2) \\ \vdots \\ \randvar{B}_1(\SSpt,J_1) \end{Bmatrix}, \quad
{\bblambda_{2}} = \begin{Bmatrix} \randvar{B}_2(\SSpt,1,1) \\ \randvar{B}_2(\SSpt,2,2) \\ \vdots \\ \randvar{B}_2(\SSpt,J_2,J_2) \end{Bmatrix}, \quad
{\bblambda_{3}} = \begin{Bmatrix} \randvar{B}_3(\SSpt,1,1,1) \\ \randvar{B}_3(\SSpt,2,2,2) \\ \vdots \\ \randvar{B}_3(\SSpt,J_3,J_3,J_3) \end{Bmatrix}, \quad
{\bblambda_{nl}} = \begin{Bmatrix} {\bblambda_{2}} \\ {\bblambda_{3}} \end{Bmatrix}.
\end{eqnarray}
\end{footnotesize}

Additionally, the index $m$ is used to represent the four different indicators calculated. So, the general notation $\bblambda_{m}$ is used from here, with $m = 1,\, 2,\, 3,$ or $nl$, depending on the number of the kernel considered in the analysis. The damage detection can be summarized as follows

\begin{enumerate}
	\item Identification of the multiple Volterra models in the reference condition, through the stochastic version of the Volterra series;
	\item Construction of the indexes based on kernels coefficients, in the reference condition ($\bblambda_{m}$);
	\item Identification of a new Volterra model in an unknown condition;
	\item Construction of the indexes based on kernels coefficients, in an unknown condition ($\bm{\lambda}_{m}$);
	\item Comparison between the coefficients estimated in unknown and reference conditions, if the new model belongs to the set of reference models, the structure is classified as healthy.  
\end{enumerate} 

The comparison between new and reference indexes will be done based on the novelty detection or outliers analysis in multivariate data, using Mahalanobis squared distance, described further.

\subsubsection{Damage detection based on kernels contribution}
\label{Damage detection based on Kernels contribution}
With the aim of comparison between the results, a different approach is proposed, based on the contribution of the Volterra kernels identified, to the total response. In this situation, the convolution between the input signal filtered by the Kautz functions and the kernels expanded in the orthonormal basis is considered, giving higher importance to the random Kautz functions. The main advantage of Volterra series approach is the capability of separate the model response in linear and nonlinear contributions, through the kernels estimated
\begin{eqnarray}
y_1(k)\approx \sum_{i_{1}=1}^{J_{1}} \mathcal{B}_1\left(i_{1}\right) \, l_{i_{1}}(k)\, ,\\
y_2(k)\approx \sum_{i_{1}=1}^{J_{2}}\sum_{i_{2}=1}^{J_{2}} \mathcal{B}_2\left(i_{1},i_{2}\right) \, l_{i_{1}}(k) \, l_{i_{2}}(k)\, , \\
y_3(k)\approx \sum_{i_{1}=1}^{J_{3}} \sum_{i_{2}=1}^{J_{3}} \sum_{i_{3}=1}^{J_{3}} \mathcal{B}_3\left(i_{1},i_{2},i_{3}\right) \, l_{i_{1}}(k) \, l_{i_{2}}(k) \, l_{i_{3}}(k)\, , \\
y_{nl}(k) = y_2(k) + y_3(k)\, .
\end{eqnarray}
\noindent
where $y_1(k)$, $y_2(k)$, $y_3(k)$ and $y_{nl}(k)$ are the linear, quadratic, cubic and nonlinear contributions, respectively. Again, considering the stochastic version of the Volterra series and the Monte Carlo simulations, multiple models are estimated in the reference condition and each vector becomes a random process
\begin{eqnarray}
\randvar{y}_1(\SSpt,k)\approx \sum_{i_{1}=1}^{J_{1}} \randvar{B}_1\left(\SSpt,i_{1}\right) \, \randvar{l}_{i_{1}}(\SSpt,k)\, ,\\
\randvar{y}_2(\SSpt,k)\approx \sum_{i_{1}=1}^{J_{2}}\sum_{i_{2}=1}^{J_{2}} \randvar{B}_2\left(\SSpt,i_{1},i_{2}\right) \, \randvar{l}_{i_{1}}(\SSpt,k) \, \randvar{l}_{i_{2}}(\SSpt,k)\, , \\
\randvar{y}_3(\SSpt,k)\approx \sum_{i_{1}=1}^{J_{3}} \sum_{i_{2}=1}^{J_{3}} \sum_{i_{3}=1}^{J_{3}} \randvar{B}_3\left(\SSpt,i_{1},i_{2},i_{3}\right) \, \randvar{l}_{i_{1}}(\SSpt,k) \, \randvar{l}_{i_{2}}(\SSpt,k) \, \randvar{l}_{i_{3}}(\SSpt,k)\, , \\
\randvar{y}_{nl}(\SSpt,k) = \randvar{y}_2(\SSpt,k) + \randvar{y}_3(\SSpt,k)\, .
\end{eqnarray}

As mentioned before, the index $m$ is used to represent the five different indexes calculated. So, the general notation $\randvar{y}_{m}$ is used from here, with $m = 1,\, 2,\, 3,$ or $nl$. The damage detection is very similar to the methodology proposed using kernels coefficients and can be summarized as follows
\begin{enumerate}
	\item Identification of the multiple Volterra models in the reference condition, through the stochastic version of the Volterra series;
	\item Calculation of the kernels contribution to the total model response, in the reference condition ($\randvar{y}_{m}$);
	\item Identification of the Volterra model in an unknown condition;
	\item Calculation of the kernels contribution to the total model response, in an unknown condition ($y_{m}$);
	\item Comparison between the contributions estimated in unknown and reference condition,  if the new model belongs to the set of reference models, the structure is classified as healthy.  
\end{enumerate} 

The difference here is that using the kernels contribution, the Kautz functions have an influence on the process, whereas using only the coefficients, they do not. Based on that, the variations in the frequency of oscillation and the damping coefficients are captured by the Kautz parameters, this version of the approach is sensitive to these variabilities. The damage detection is also done based on the novelty detection, considering multivariate data. 

\subsubsection{Novelty detection}
\label{Novelty detection}
The indexes proposed based on Volterra models are described as random processes in the reference condition. This methodology combined with Monte Carlo simulations allows the generation of a set of reference models that can be used as monitoring parameters, as described before. So, the damage detection can be done based on the novelty detection or outliers analysis, considering multivariate data. Let us consider the set of indexes in the reference condition ($\bblambda_{m}$ or $\randvar{y}_m$), the classical Mahalanobis squared distance can be calculated \cite{WORDEN2000647}
\begin{eqnarray}
\randvar{D}_{m}(\SSpt) = [\bblambda_{m}(\SSpt,i) - \mean{\bblambda_{m}}]^T \, \bm{\Sigma}^{\mbox{-}1} \, [\bblambda_{m}(\SSpt,i) - \mean{\bblambda_{m}}]  
\end{eqnarray}
\noindent
or
\begin{eqnarray}
\randvar{D}_{m}(\SSpt) = [\randvar{y}_m(\SSpt,k) - \mean{\randvar{y}_m}]^T \, \bm{\Sigma}^{\mbox{-}1} \, [\randvar{y}_m(\SSpt,k) - \mean{\randvar{y}_m}]
\end{eqnarray}  
\noindent
where $\randvar{D}_{m}(\SSpt)$ is the Mahalanobis squared distance in the reference condition, $\bm{\Sigma}$ is the covariance matrix and $\mean{(.)}$ is the mean operator. Note that, in the reference condition, $\randvar{D}_{m}(\SSpt)$ is a random variable, with a correspondent probability density function (PDF). So, its PDF can be empirically estimated considering the Kernel Density Estimation approach in order to establish a threshold value to the distribution \cite{silverman1986density,WORDEN2003323}
\begin{equation}
\hat{p}_{\randvar{D}_{m}}(d_m) = \frac{1}{Nh} \sum_{i = 1}^{N} K \left(\frac{d_{m} - d_{m,i}}{h}\right) \, ,
\label{PDFlambda}
\end{equation}
\noindent 
where $\hat{p}_{\randvar{D}_{m}}(d_m)$ is an approximation of $p_{\randvar{D}_{m}}(d_m)$ that is the true density of $\randvar{D}_{m}(\SSpt)$, $N$ is the total number of realizations of the random parameter, $d_{m,i}$ is the $i$th realization of the random variable $\randvar{D}_m(\SSpt)$, $K$ represents the kernel of the transformation, in this work the Gaussian kernel is used, and $h$ is the smoothing parameter that controls the width of the Gaussian kernel chosen. The main difficulty in the Kernel Density Estimation approach is the choice of the optimal value to the smoothing parameter and, in this work, the cross-validation was used to set the better value to $h$ \cite{Bowman1997}. The Kernel Density Estimation is used, because the distribution of $\randvar{D}_{m}(\SSpt)$ is assumed as unknown a priori and the large number of Monte Carlo simulations, provides sufficient samples to estimate its density using this approach. 

\begin{figure}[!htb]
	\begin{center}
		\includegraphics[scale=0.5]{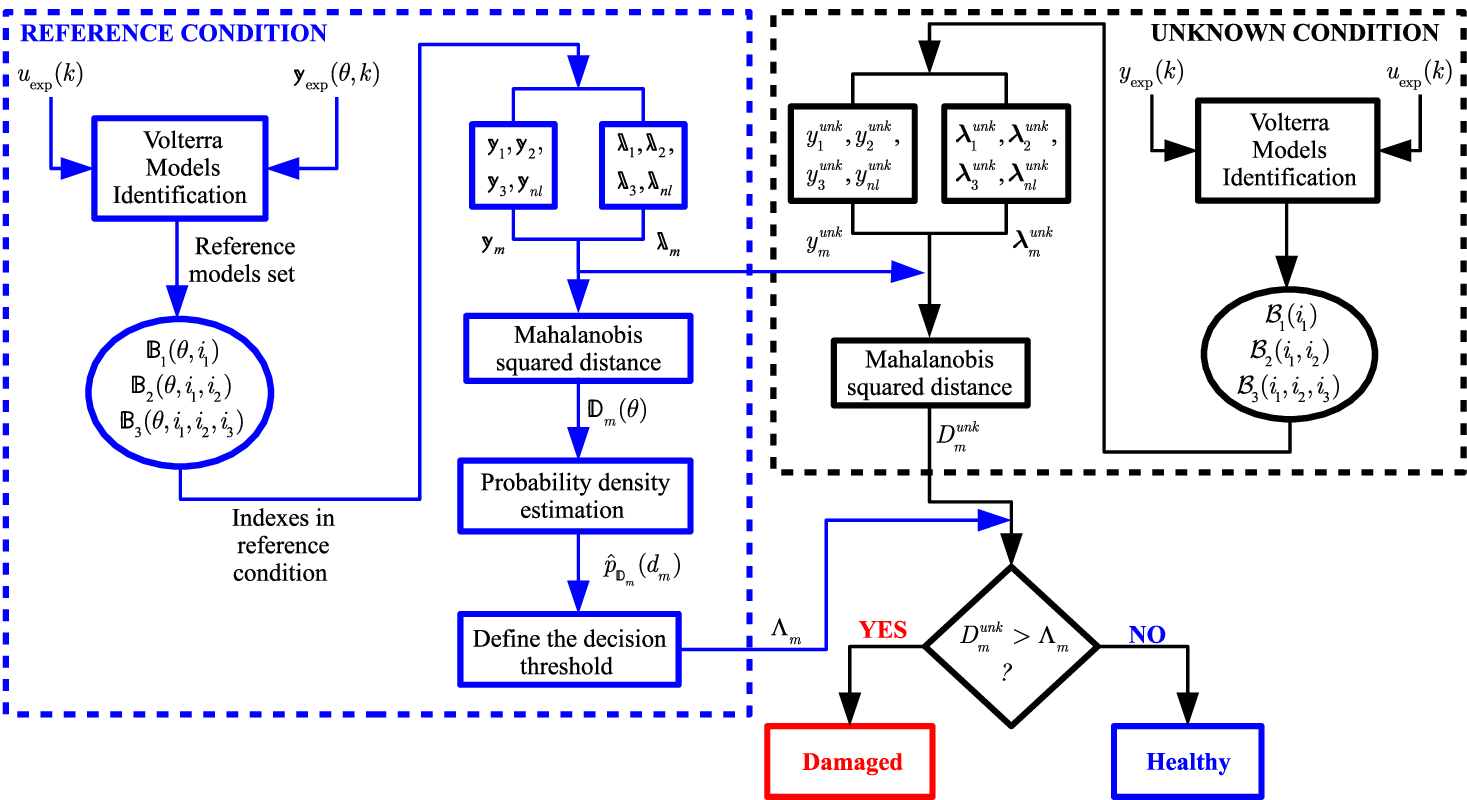}
		\caption{Description of the damage detection approach based on stochastic Volterra series.}
		\label{fluxodamage}
	\end{center}
\end{figure}

Then, with the density estimated $\hat{p}_{\randvar{D}_{m}}(d_m)$  it is possible to establish a threshold value for the distribution, which can be used in the damage detection procedure. The threshold value can be defined as \cite{MarcRebillat2016}
\begin{equation}
\Lambda_m = \{d_m \, \,\mbox{such that} \int_{d_m}^{+\infty}  p_{\randvar{D}_{m}}(d_m)\, \mbox{d}d_{m} = \beta\} \, ,
\label{EqThreshold}
\end{equation}
\noindent 
where $\Lambda_m$ represents the threshold value and $\beta$ is the sensitivity of the approach or probability of false alarms considered. The better value of $\beta$ to be chosen it depends on the level of security/confidence that each operational application demands. Now, a new model is identified in an unknown condition (healthy or damaged). The indexes are computed considering the new model ($\lambda_{m}^{unk}$ or $y_m^{unk}$) and compared with the stochastic model (set of reference models), through the Mahalanobis squared distance
\begin{eqnarray}
D_{m}^{unk} = [\lambda_{m}^{unk} - \mean{\bblambda_{m}}]^T \, \bm{\Sigma}^{\mbox{-}1} \, [\lambda_{m}^{unk} - \mean{\bblambda_{m}}]  
\end{eqnarray}
\noindent
or
\begin{eqnarray}
D_{m}^{unk} = [{y}_m^{unk} - \mean{\randvar{y}_m}]^T \, \bm{\Sigma}^{\mbox{-}1} \, [{y}_m^{unk} - \mean{\randvar{y}_m}]
\end{eqnarray} 
\noindent
where $D_{m}^{unk}$ is the Mahalanobis squared distance in the unknown condition. Finally, the hypothesis test can be applied to determine if the system is in healthy or damaged condition 
\begin{equation}
\left\{\begin{array}{c}
H_0 :D_{m}^{unk}\leq\Lambda_m\, ,\\
H_1 :D_{m}^{unk}>\Lambda_m\, ,
\end{array}\right.
\end{equation}
\noindent where the null hypothesis $H_0$ represents the healthy condition and $H_1$ the damaged. The methodology used to detect damage is summarized in the flowchart showed in fig. \ref{fluxodamage}, considering the two proposed approaches.

\FloatBarrier
%
\section{Modeling a breathing crack in an uncertain nonlinear beam}
\label{Nonlinear System of Interesting}
This section describes the studied nonlinear system that can be approximated by a nonlinear single-degree-of-freedom (SDOF) model, a Duffing oscillator \cite{kovacic2011duffing}. The deterministic reference system is presented, then, the simulated damage used, a breathing cracked model, is shown. Finally, the upgrade to the stochastic version is done with the necessary information about the uncertainties considered.

\subsection{Deterministic system in reference condition}
\begin{figure}[!htb]
	\begin{center}
		\includegraphics[scale=0.5]{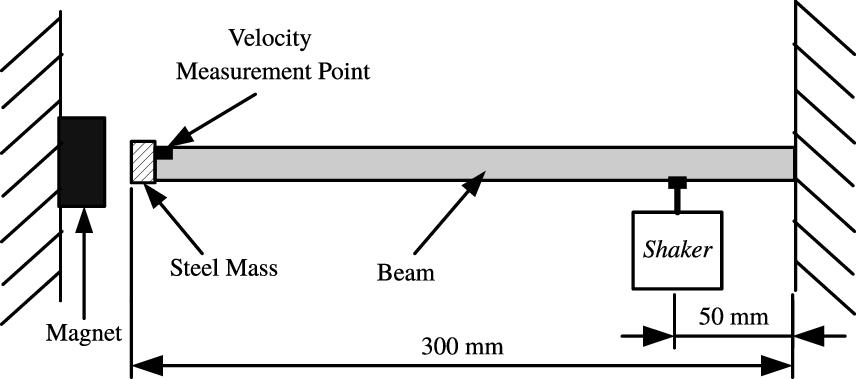}
		\caption{Nonlinear system simulated in the reference condition.}
		\label{NonlinearBeam}
	\end{center}
\end{figure}

In this work, a single-degree-of-freedom (SDOF) model is used to simulate a real system, composed by a clamped-free aluminum beam with a steel mass positioned in its free extremity, to generate a nonlinear interaction with a magnet (fig. \ref{NonlinearBeam}). The objective of the setup used is to emulate a hardening nonlinear system, to study the performance of the damage detection approach proposed. In all simulations, the input and output signals considered are the force applied by the shaker and the velocity measured near to the free extremity of the beam, respectively. The real system presents nonlinear behavior only for large displacements and its dynamic behavior can be well approximated, for a limited frequency range, by its first mode shape that can be represented as a Duffing oscillator \cite{kovacic2011duffing}  
\begin{equation}
m \, \ddot{x}(t) + c \,  \dot{x}(t) + k_1 \,  x(t) +  k_2 \, x(t)^{2} +  k_3 \, x(t)^{3} = U(t),
\label{DuffingEq1}
\end{equation}
\noindent
where $m$ is the system equivalent mass in [kg], $c$ is the damping coefficient in [Ns/m], $k_1$ is the linear stiffness in [N/m], $k_2$ is the quadratic stiffness in [N/m$^2$], $k_3$ is the cubic stiffness in [N/m$^3$], and $U(t)$ is the external force in [N].  The displacement, velocity and acceleration in the free extremity of the beam are, respectively, represented by $x(t)$, $\dot{x}(t)$ and $\ddot{x}(t)$. The quadratic stiffness was added to emulate the interaction between shaker and beam, to ensure a more realistic description of the system behavior, whereas the cubic stiffness describes the interaction between beam and magnet.

The deterministic values of the Duffing oscillator parameters were estimated and are shown in Table. \ref{DuffingParameters}. The estimation of the parameters was performed considering modal analysis and the restoring force surface (RFS) method, considering experimental data measured. The description of the procedure used to approximate the real system considered by the Duffing oscillator model it is not shown here because it is not the focus of this work.  

\begin{table}[!htb]
	\caption{Duffing oscillator parameters.}
	\label{DuffingParameters}
	\begin{center}
		\begin{tabular}{| c || c | }
			\hline
			\textbf{Parameter} & \textbf{Deterministic value} \\
			\hline
			$m$ [kg] & 0.26  \\
			$c$ [Ns/m] & 1.36 \\
			$k_1$ [N/m] & $5.49 \times 10^3$ \\
			$k_2$ [N/m$^2$] & $3.24 \times 10^4$  \\
			$k_3$ [N/m$^3$] & $4.68 \times 10^7$  \\
			\hline
		\end{tabular}
	\end{center}
\end{table}  

In order to illustrate the nonlinear behavior of the system in the reference condition (Healthy), some tests were carried out. In all the tests, it was used a sampling frequency of 512 Hz, 2048 samples and it was considered two levels of the input signal (0.1 N - Low level and 1 N - High level) in order to study the linear and nonlinear behavior of the system. The system was excited by a chirp signal varying the excitation frequency from 15 to 30 Hz in 4 seconds.  

\begin{figure}[!htb]
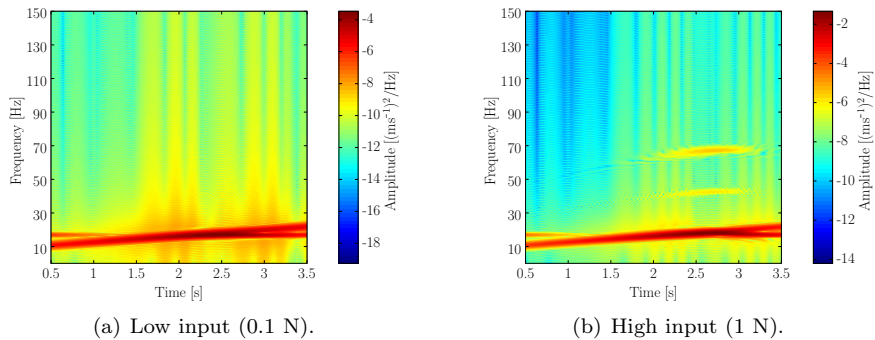

	\begin{center}
		\subfigure[Low input (0.1 N).]{\scalebox{0.27}{\huge \input{Figures/NonlinearBehavior/SpecRespLow.tex}}} \qquad
		\subfigure[High input (1 N).]{\scalebox{0.27}{\huge \input{Figures/NonlinearBehavior/SpecRespHigh.tex}}} 
		\caption{Time-frequency diagram of the response for different levels of input amplitude, considering the reference condition (Healthy).}
		\label{Spec}
	\end{center}
\end{figure}

Figure \ref{Spec} shows the time-frequency diagram of the system response signals, considering the low and high level of the input amplitude. The nonlinear effect can be observed with the appearance of multiple harmonics (2 and 3 times the fundamental harmonic) when a high level of input is applied. These results confirm that the system has linear behavior to the low level of input signal and nonlinear to the high level of input, in the reference condition. The presence of multiple harmonics is related to the nonlinear stiffnesses of the Duffing oscillator motion equation. These results show, clearly, the nonlinear behavior of the system for large displacements (condition reached considering 1 N as excitation amplitude).

\subsection{Damage simulated $-$ A breathing crack model}
\label{Breathing Crack Model}
\begin{figure}[!htb]
	\begin{center}
		\includegraphics[scale=0.5]{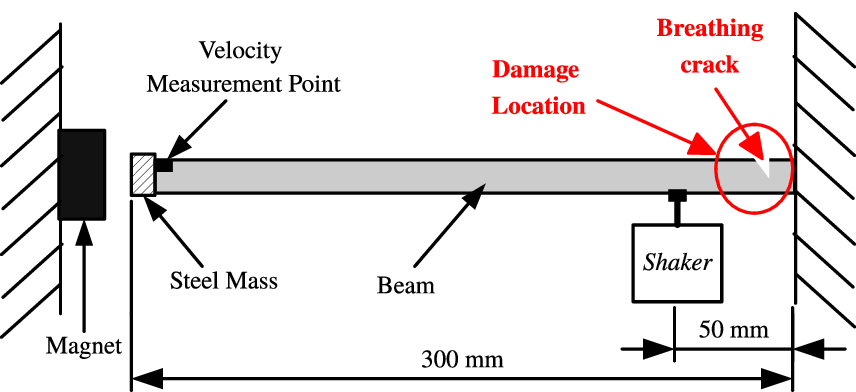}
		\caption{Nonlinear system simulated in the damaged condition.}
		\label{NonlinearCrackedBeam}
	\end{center}
\end{figure}

The methodology showed in section \ref{The Sotochastic Volterra Series Approach for Damage Detection} proposes to detect damages, or structural variations, in nonlinear systems, considering data variation. In this sense, a breathing crack model is used to emulate the presence of a damage in the reference studied system. Figure \ref{NonlinearCrackedBeam} shows the presence of a crack in the system of interesting.   

The objective is to detect the presence of the crack in the beam, considering that the system has nonlinear behavior before (Duffing oscillator) and after (breathing crack) the damage occurrence, taking into account the data variation. In other words, the nonlinear characteristic of the damage has different nature of the nonlinear behavior of the system in the reference condition. A lot of papers, over the years, have shown the approximation of the breathing crack phenomenon by a single-degree-of-freedom (SDOF) model, as a bilinear oscillator \cite{CHATI1997249,ANDREAUS2007566,Chatterjee20103325,Prawin2018}. Combining the reference model (Duffing oscillator) with the cracked beam model, the system behavior can be approximated by   
\begin{eqnarray}
m \, \ddot{x}(t) + c \,  \dot{x}(t) + \mathcal{F}(x(t),t) +  k_2 \, x(t)^{2} +  k_3 \, x(t)^{3} = U(t),
\label{SystemEquation} 
\end{eqnarray} 
\noindent where $\mathcal{F}(x(t),t)$ is the bilinear force
\begin{equation}
\mathcal{F}(x(t),t)=
\begin{cases}
k_1 \, x(t), & \mbox{if} \, \, \, \, x(t)\geq 0 \\
\alpha \, k_1 \, x(t), & \mbox{if} \, \, \, \, x(t)< 0
\end{cases}
\nonumber
\end{equation}
\noindent where $0<\alpha<1$ represents the crack severity. If $\alpha = 1$ the system is in the reference condition (without crack). The breathing crack phenomenon generates a nonlinear behavior with different nature of the cubic effect of the reference model but similar to the quadratic effect of $k_2$. This situation makes the application of damage detection procedures difficult. The applied approach has to be able to detect the nonlinear damage without confusions with the nonlinear behavior of the reference condition. This is a great improvement compared with other authors like Peng et al. (2007) \cite{PENG2007777}, Surace et al. (2011) \cite{Surace2011} and Prawin et al. (2018) \cite{Prawin2018}, for example, that have used the Volterra series to detect damage, considering the linear behavior in reference condition and the nonlinear effects as a consequence of the crack occurrence. 

\begin{figure}[!htb]
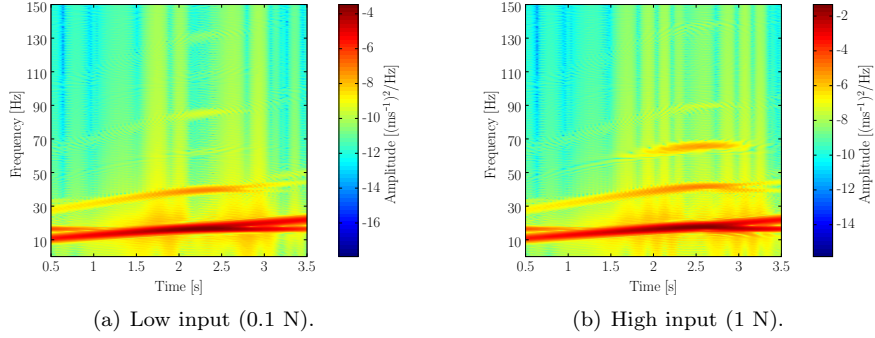

	\begin{center}
		\subfigure[Low input (0.1 N).]{\scalebox{0.27}{\huge \input{Figures/NonlinearBehavior/SpecRespLowCrack.tex}}} \qquad
		\subfigure[High input (1 N).]{\scalebox{0.27}{\huge \input{Figures/NonlinearBehavior/SpecRespHighCrack.tex}}} 
		\caption{Time-frequency diagram of the response for different levels of input amplitude, considering the crack severity $\alpha = 0.9$.}
		\label{SpecCrack}
	\end{center}
\end{figure}

In the presence of the breathing crack, the nonlinear behavior does not depend on the amplitude of the response and can be observed for all excitation amplitudes. Figure \ref{SpecCrack} shows the time-frequency diagram of the system response signals to the low and high level of amplitude excitation, considering the presence of a crack with $\alpha = 0.9$. It can be observed, that to the low and high level of input amplitude, the appearance of quadratic harmonics, mainly of the second order. This behavior brings closer the breathing crack phenomenon by a SDOF model and it is a consequence of the bilinear characteristic of the response. 

\subsection{Stochastic version of the nonlinear system}
\label{Stochastic Model}
It is known that mechanical systems and structures are subjected to uncertainties that have to be considered in the processes of models identification and damage detection procedures, as it was exemplified in section \ref{Introduction}. So, taking the uncertainties into account is essential to make robust predictions \cite{CunhaJr2017252}. The data uncertainties considered are related with environmental or operating conditions, temperature and humidity changes, sensor bonding conditions, that can generate variations in the fundamental frequencies and damping ratios of the structures and systems, making difficult the application of damage detection procedures \cite{Sohn539}. The main idea is to simulate these variabilities in order to reproduce, in a more realistic way, the difficulties involved in the damage detection procedures. In this context, it is not considered the simple noise addition in the measurements, but the variations in the system dynamic behavior.      

In order to simulate the presence of uncertainties in the system and to test the performance of the stochastic identification and damage detection approach proposed in section \ref{The Sotochastic Volterra Series Approach for Damage Detection}, the linear stiffness ($k_1$) and damping coefficient ($c$) of Eq. (\ref{SystemEquation}) are considered as random variables $\SSpt \in \SS \mapsto \randvar{k}_1 (\SSpt) \in \R$, $\SSpt \in \SS \mapsto \randvar{c} (\SSpt) \in \R$. The variation of $k_1$ and $c$ reflects directly in the system dynamics, since these parameters influence the fundamental frequency of oscillation and the damping ratio of the system.  So, the Eq. (\ref{SystemEquation}) can be rewritten as
\begin{equation}
m \, \ddot{\randvar{x}}(\SSpt,t) + \randvar{c}(\SSpt) \,  \dot{\randvar{x}}(\SSpt,t) + \randvar{F}(\randvar{x}(\SSpt,t),t) +  k_2 \, \randvar{x}(\SSpt,t)^{2} +  k_3 \, \randvar{x}(\SSpt,t)^{3} = U(t),
\label{DuffingEq2}
\end{equation}
\noindent where the bilinear random force $\randvar{F}(\randvar{x}(\SSpt,t),t)$ can be described as
\begin{equation}
\randvar{F}(\randvar{x}(\SSpt,t),t)=
\begin{cases}
\randvar{k}_1(\SSpt) \,  \randvar{x}(\SSpt,t), & \mbox{if} \, \, \, \, \randvar{x}(\SSpt,t)\geq 0 \\
\alpha \, \randvar{k}_1(\SSpt) \,  \randvar{x}(\SSpt,t), & \mbox{if} \, \, \, \, \randvar{x}(\SSpt,t)< 0
\end{cases}
\nonumber
\end{equation}

\noindent and the random processes $(\SSpt,t) \in \SS \times \R \mapsto \randvar{x}(\SSpt,t)$, 
$(\SSpt,t) \in \SS \times \R \mapsto \dot{\randvar{x}}(\SSpt,t)$, and 
$(\SSpt,t) \in \SS \times \R \mapsto \ddot{\randvar{x}}(\SSpt,t)$, represent, respectively, the displacement, velocity and acceleration in the beam free extremity. 

\begin{figure}[!htb]
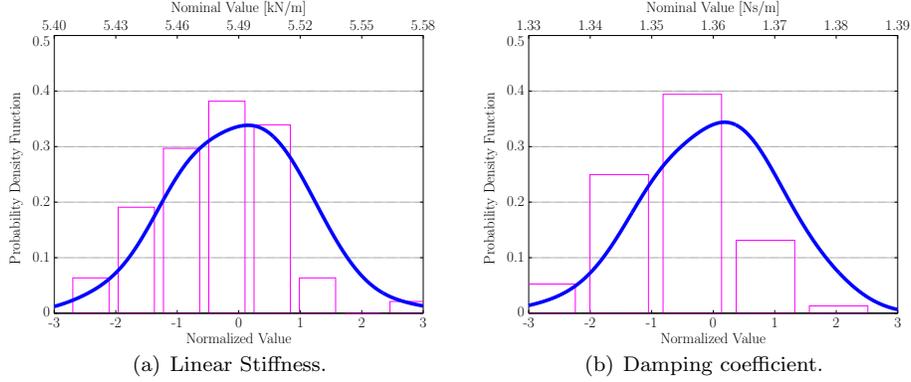

	\begin{center}
		\vspace{0.3cm}
		\subfigure[Linear Stiffness.]{\scalebox{0.27}{\huge \input{Figures/StochasticModel/Distribution_k.tex}}} \qquad
		\subfigure[Damping coefficient.]{\scalebox{0.27}{\huge \input{Figures/StochasticModel/Distribution_c.tex}}} 
		\caption{Distribution of the random parameters $\randvar{k}_1$ and $\randvar{c}$.}
		\label{MandC_Dist}
	\end{center}
\end{figure}

To construct the probabilistic model, the PDFs of the random parameters were determined through the maximum entropy principle \cite{PhysRev,cunhajr2017}. Considering that the linear stiffness and the damping coefficient can not be negative, we assumed the interval $(0,\infty)$ as the support of these random variables. Also, it was considered that the expected value of $\randvar{k_1}$ and $\randvar{c}$ are known real numbers $\mu_{\randvar{k_1}}$ and $\mu_{\randvar{c}}$. For technical reasons, see Soize (2017) \cite{Soize2017} for details, we also supposed that the expected value of $\ln{\randvar{k_1}}$ and $\ln{\randvar{c}}$ are finite. Using these conditions as known information, such as done in Cunha and Sampaio (2015) \cite{CunhaJr2015809}, the principle of maximum entropy says that the probability density function (PDF) of these random variables is given by
\begin{equation}
\pdf{\randvar{Z}}(z) = \mathbb{1}_{(0,\infty)} \frac{1}{\mu_{\randvar{Z}}} \left( \frac{1}{\delta_{\randvar{Z}}^2} \right)^{ \left( \frac{1}{\delta_{\randvar{Z}}^2} \right) } \frac{1}{\Gamma(1/\delta_{\randvar{Z}}^2)} \left( \frac{z}{\mu_{\randvar{Z}}} \right)^{ \left( \frac{1}{\delta_{\randvar{Z}}^2}-1 \right) } \exp\left( - \frac{z}{\delta_{\randvar{Z}}^2 \, \mu_{\randvar{Z}}} \right)
\label{gamma_distrib}
\end{equation}

\noindent where $\randvar{Z}$ represents the random parameter ($\randvar{k}_1$ or $\randvar{c}$), $\mu_{\randvar{Z}}$ is the mean value, $\delta_{\randvar{Z}}$ is the dispersion, $\Gamma$ indicates the gamma function, and $\mathbb{1}_{(0,\infty)}$ denotes the indicator function of the interval $(0,\infty)$. This PDF corresponds to a gamma distribution. The stochastic formulation was made assuming, for the lack of better knowledge, that $\randvar{k}_1$ and $\randvar{c}$ are independent random variables. To compute the propagation of the uncertainties of the random parameters $\randvar{k}_1$ and $\randvar{c}$ through the model, the Monte Carlo (MC) method \cite{rubinstein2016simulation} was employed, using 64 samples of each parameter, combined to generate a total of 2048 MC realizations of the deterministic system. The simulations were performed considering $\mu_{\mathbb{k}_1} = 5.49 \times 10^3$ [N/m], $\delta_{\mathbb{k}_1} = 0.01$, $\mu_{\mathbb{c}} = 1.36$ [Ns/m], $\delta_{\mathbb{c}} = 0.01$. These values of dispersion were chosen to generate satisfactory variation in the system response. Figure \ref{MandC_Dist} shows the probability densities of the random parameters used in the MC simulations. The variation of the random parameters makes the system response varies too and the data variation will be considered in the model identification and damage detection procedures. 

\begin{figure}[!htb]
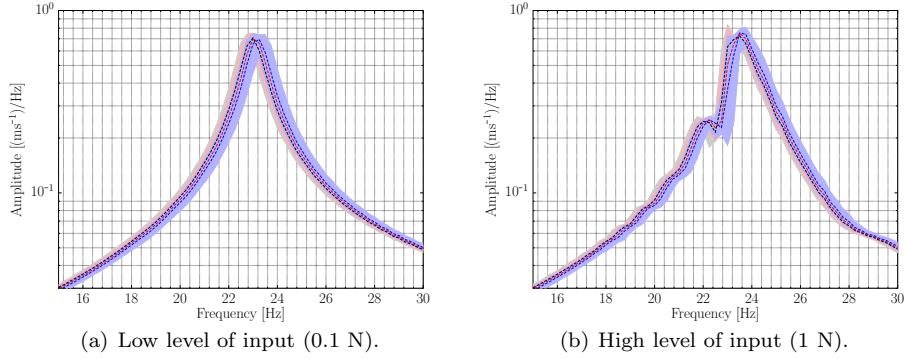

	\begin{center}
		\subfigure[Low level of input (0.1 N).]{\scalebox{0.27}{\huge \input{Figures/StochasticModel/FRFsLow.tex}}} \qquad
		\subfigure[High level of input (1 N).]{\scalebox{0.27}{\huge \input{Figures/StochasticModel/FRFsHigh.tex}}} 
		\caption{FRFs considering different structural conditions and 2 levels of input amplitude. \colorbox{test}{\textcolor{test}{B}} represents $\alpha = 1$ (reference condition), \colorbox{gray}{\textcolor{gray}{B}} $\alpha = 0.98$ (damaged condition) and \colorbox{test2}{\textcolor{test2}{B}} $\alpha = 0.96$ (damaged condition).}
		\label{FRFs}
	\end{center}
\end{figure}

To exemplify the difficulty to detect damage in this situation, fig.~\ref{FRFs} shows the 99\% confidence bands of the system Frequency Response Function (FRF), considering 3 different structural conditions. It is not possible to distinguish the damaged conditions of the reference state, considering the variation of the response, since it can be observed an overlap of the envelopes. So, the use of modal parameters or the direct system response to detect damage in this situation is not recommended, because only conditions of high severity damage can be detected with probabilistic confidence. The propose is to use the nonlinear behavior of the damage as an indicator of the structural variation. However, the nonlinear behavior in the reference condition has to be considered to avoid the confusion with damage.

Next section shows the application of the methodology proposed to detect damages, based on stochastic Volterra series, in the cracked beam model described, considering the initially nonlinear behavior and data variation.

\section{Damage detection procedure}
\label{Numerical Results}
This section shows the principal results obtained with the application of the damage detection approach proposed, based on the stochastic Volterra series, in the nonlinear system described in section \ref{Nonlinear System of Interesting}, considering the nonlinear behavior of the system in the reference condition and the data variation. In all simulations, it was used a sampling frequency of 512 Hz and 2048 samples. The Monte Carlo simulations were performed considering 2048 realizations, to ensure the method convergence. The response of the system described in section \ref{Nonlinear System of Interesting} was simulated considering different structural conditions, i. e.,  $\alpha = 1.00,\, 0.98,\, 0.96,\, 0.94,\, 0.92,\, 0.90,\, 0.88,\, 0.86$ (2 reference and 7 damage conditions) and the simulated data variation. The simulations regarding the reference condition were repeated twice, 2048 realizations were used as training data and 2048 as test data. To better emulate a real application, it was added to the system response, in all situations simulated, a Gaussian noise in order to have a signal to noise ratio (SNR) of 30 dB.

\subsection{Stochastic Volterra models identification}
\label{Stochastic Volterra model identification}
The identification of the stochastic Volterra model was done considering 2048 realizations, for each structural condition, obtained through the simulations performed. Only the first three kernels were used to identify the system, once its nonlinear characteristic in the healthy and damaged condition can be approximated by  the quadratic and cubic term. First of all, the Kautz parameters were parametrized in terms of the natural frequency ($\bbomega_{n} (\SSpt)$) and the damping ratio ($\bbzeta (\SSpt)$) of the equivalent linear system. As the uncertainties considered in this work influence on the modal parameters, they vary in each realization. So, they were estimated to each response realization and the optimization process described in section \ref{The stochastic Kautz parameters} was used to define the Kautz parameters, taking into account the relations described in Eqs. (\ref{Kautzrelationship}). Table \ref{KautzParam} shows the values of the factors $p_1$, $p_2$, $p_3$ and $p_4$ found for the high order kernels, considering the different damage severities. Changes are observed in the position of the poles as the crack increases, mainly related with the second order kernel, because of the quadratic effect of the crack. The number of functions were chosen as done in Shiki et al. (2017) \cite{BruceSHMjournal}, and defined as $J_1 = 2$, \textcolor{red}{$J_2 = 4$} and $J_3 = 6$.  

\begin{table}[!htb]
	\caption{Relation factors between  the Kautz and modal parameters, considering the high order Volterra kernels.}
	\label{KautzParam}
	\begin{center}
		\begin{tabular}{| c || c | c | c | c |}
			\hline
			\textbf{Crack severity} & \multicolumn{2}{c|}{ Second kernel} & \multicolumn{2}{c|}{ Third kernel}\\
			\cline{2-5}
			\textbf{($\alpha$)} & $p_1$ & $p_2$ & $p_3$ & $p_4$ \\
			\hline
			1.00 & 1.11  & 2.7 & 1.06 & 1.1\\
			0.98 & 1.07 & 2.4 & 1.06 & 1.1\\
			0.96 & 1.07 & 2.3 & 1.06 & 1.1\\
			0.94 & 1.06 & 2.2 & 1.06 & 1.1\\
			0.92 & 1.06 & 2.1 & 1.06 & 1.1\\
			0.90 & 1.05 & 2.0 & 1.06 & 1.0\\
			0.86 & 1.04 & 1.8 & 1.05 & 1.0\\
			\hline
		\end{tabular}
	\end{center}
\end{table} 

With the Kautz poles defined, the random Kautz functions can be obtained and used in the models' identification process. The identification of the Volterra models was performed considering a chirp signal as input applied to the system, varying the excitation frequency from 15 to 30 Hz (first mode region). The kernels identification was carried out in two steps, first, the linear kernel was estimated considering low level of input (0.1 N), and then, the high order kernels were estimated considering a high level of input (1 N), as done in Shiki et al. (2017) \cite{BruceSHMjournal}. The models were estimated via MC simulation, in a total of 2048 realizations for each structural condition.

\begin{figure}[!htb]
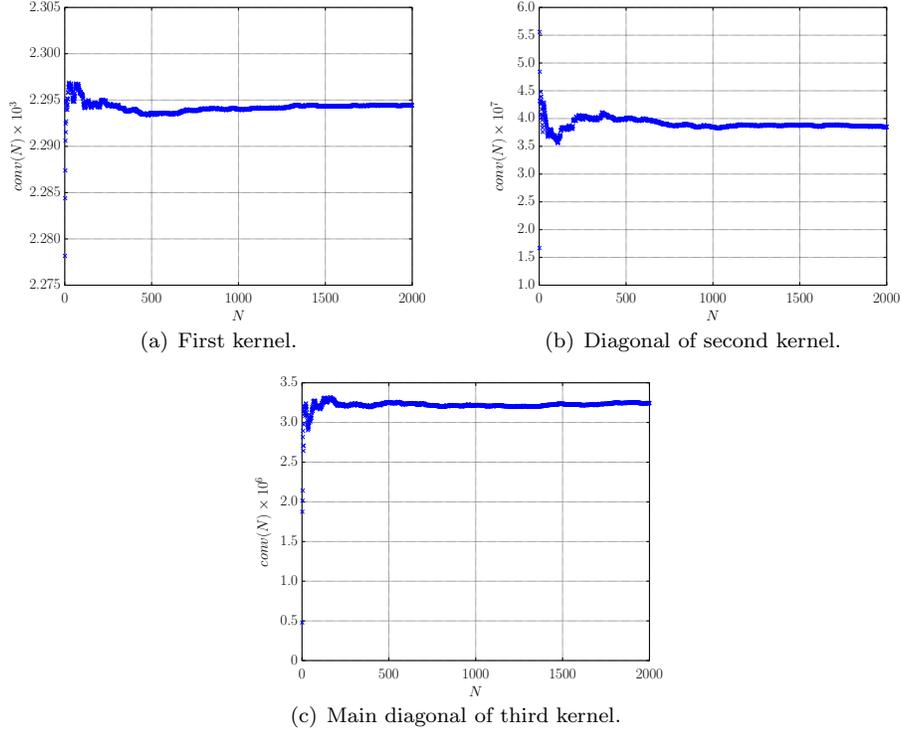

	\begin{center}
		\subfigure[First kernel.]{\scalebox{0.27}{\huge \input{Figures/MCconvergence/MCfirstKernel.tex}}} \qquad
		\subfigure[Diagonal of second kernel.]{\scalebox{0.27}{\huge \input{Figures/MCconvergence/MCsecondKernel.tex}}} \\
		\subfigure[Main diagonal of third kernel.]{\scalebox{0.27}{\huge \input{Figures/MCconvergence/MCthirdKernel.tex}}} 
		\caption{MC convergence test applied to the Volterra kernels identified in the reference condition ($\alpha = 1.00$).}
		\label{ConvKernels}
	\end{center}
\end{figure}

The convergence of the MC method is measured with the aid of a function that depends on the kernel vector or its main diagonal (high order kernels) represented in the time domain, defined by
\begin{equation}
conv(N) = \sqrt{\frac{1}{N}\sum_{n=1}^{N}\int_{t=t_0}^{t_f} || {\randvar{h}}(\SSpt_n,t) ||^2dt} \, ,
\end{equation}
\noindent
where $N$ is the number of MC simulations, $||.||$ denotes the standard Euclidean norm and $\randvar{h}(\SSpt_n,t)$ represents the $n$-th realization of the random first kernel or main diagonal of high order kernels, in the time domain. For further details, the reader is encouraged to see Soize (2005) \cite{Soize2005623}. The criterion was applied to the first, second and third kernels. Figure \ref{ConvKernels} shows the results obtained, considering the Volterra kernels identified in the reference condition. It can be seen that the MC convergence is achieved with the number of samples used. 

\subsection{Stochastic Volterra models validation}
\label{Stochastic Volterra model validation}
\begin{figure}[!htb]
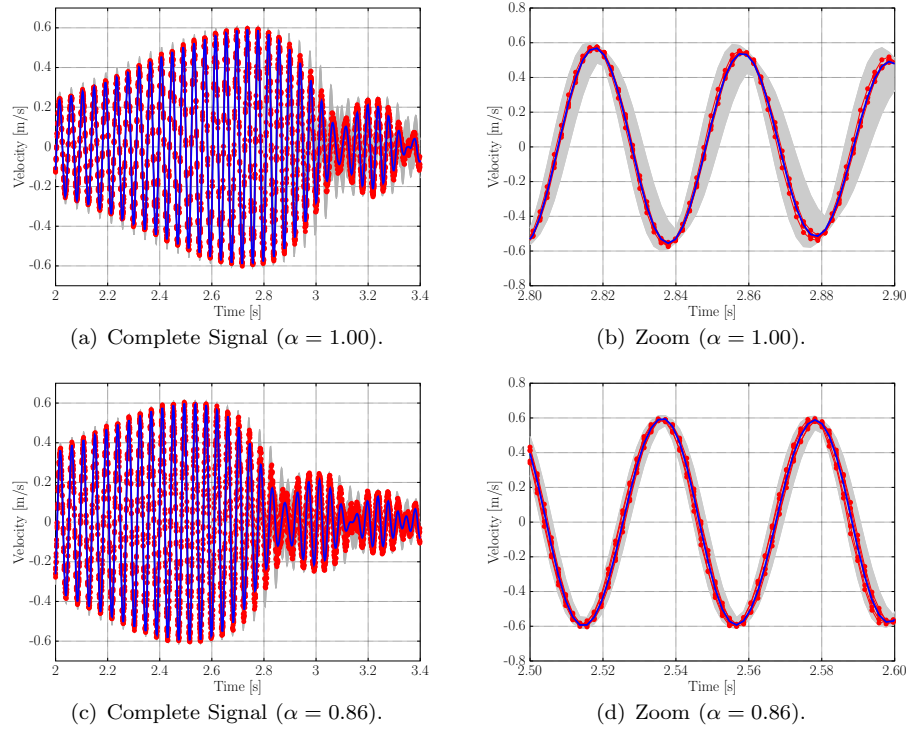

	\begin{center}
		\subfigure[Complete Signal ($\alpha = 1.00$).]{\scalebox{0.27}{\huge \input{Figures/ReferenceModel/RespHigh.tex}}} \qquad
		\subfigure[Zoom ($\alpha = 1.00$).]{\scalebox{0.27}{\huge \input{Figures/ReferenceModel/RespHigh2.tex}}} \\
		\subfigure[Complete Signal ($\alpha = 0.86$).]{\scalebox{0.27}{\huge \input{Figures/DamageModel/RespHighDamage.tex}}} \qquad
		\subfigure[Zoom ($\alpha = 0.86$).]{\scalebox{0.27}{\huge \input{Figures/DamageModel/RespHigh2Damage.tex}}}
		\caption{Stochastic Volterra model response in comparison with new simulated data, considering a high level amplitude chirp input (1 N) and two structural conditions. The gray box represents the 99\% confidence bands, \textcolor{blue}{-- --} represents the mean and \textcolor{red}{-- $\circ$ --} the simulated data.}
		\label{RespChirpHigh}
	\end{center}
\end{figure}

Before the application of the Volterra series models in the damage detection process, it is important to verify if the Volterra kernels describe the dynamical behavior of the system with certain statistical confidence. Thus, the same input used to estimate the Volterra kernels was applied to the model. Figure \ref{RespChirpHigh} shows the comparison between the 99\% confidence bands of the stochastic Volterra model and new simulated data, considering high level of input (1 N) and two structural conditions, reference ($\alpha = 1.00$) and severe damage ($\alpha = 0.86$). It can be seen that the model is able to describe the system behavior and to predict the response, even in the presence of nonlinear behavior and data variation, in both structural conditions. 

\begin{figure}[!htb]
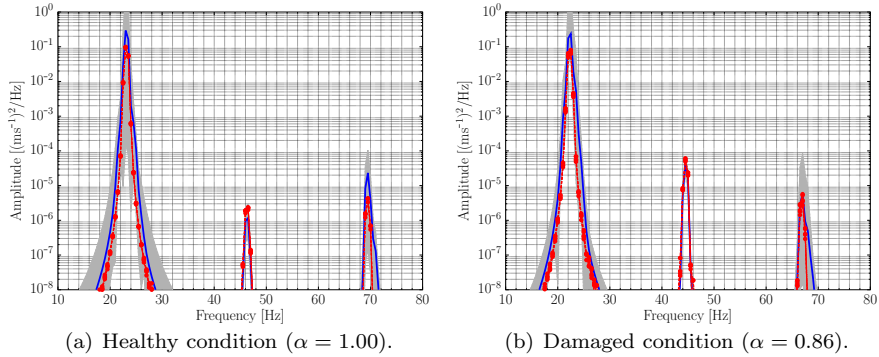

	\begin{center}
		\subfigure[Healthy condition ($\alpha = 1.00$).]{\scalebox{0.27}{\huge \input{Figures/ReferenceModel/PsdHigh.tex}}} \quad
		\subfigure[Damaged condition ($\alpha = 0.86$).]{\scalebox{0.27}{\huge \input{Figures/DamageModel/PsdHighDamage.tex}}} 
		\caption{Stochastic Volterra model response in comparison with new simulated data, in frequency domain, considering a single tone sine input and two structural conditions. The gray box represents the 99\% confidence bands, \textcolor{blue}{-- --} represents the mean and \textcolor{red}{-- $\circ$ --} the simulated data.}
		\label{RespSine}
	\end{center}
\end{figure}

Then, a signal with different nature was applied to the system to validate the stochastic Volterra model in the frequency domain. A single tone sine was applied with a high level of input (1 N) and frequency close to the equivalent linear system natural frequency ($\approx 23$ Hz), in order to amplify the nonlinear behavior. Figure \ref{RespSine} shows the results obtained in the frequency domain, to help the visualization of the harmonic components,  with 99\% of statistical confidence in comparison with new simulated data. Again, two structural conditions are shown, healthy ($\alpha = 1.00$) and severe damage ($\alpha = 0.86$). It can be seen that the model predicts the system behavior in healthy (fig. \ref{RespSine}a) and damaged (fig. \ref{RespSine}b) conditions. All frequency components present in the signal can be described by the stochastic Volterra model. The large dispersion observed is caused by the number of Kautz functions used on the description of the third order Volterra kernel, that amplifies the propagation of the uncertainties into the model, but a lower number of functions does not do the model able to describe the nonlinear behavior of the studied system. With the stochastic Volterra model validated, the stochastic response can be used to detect damage in the system as shown in the next section.

\begin{figure}[!htb]
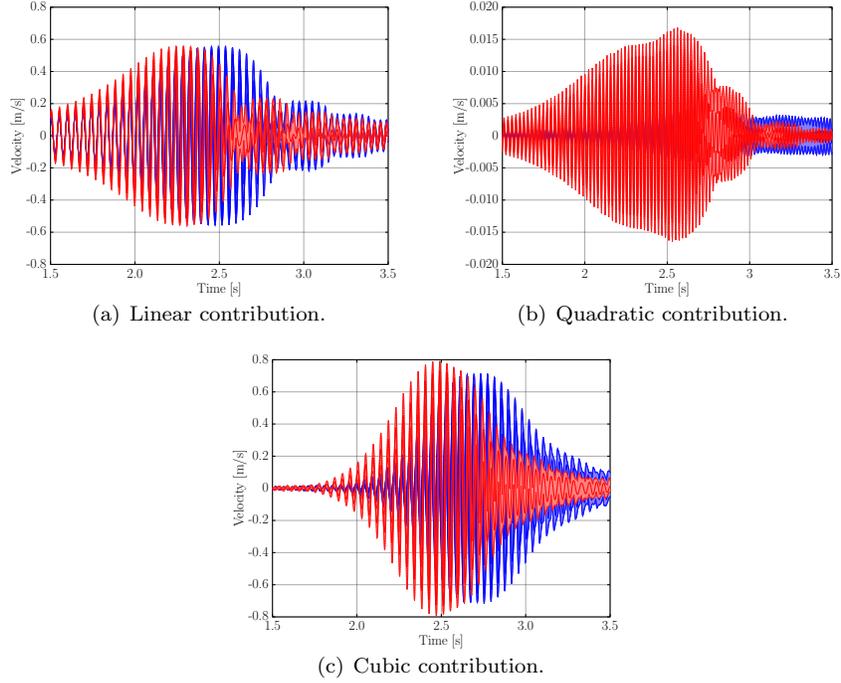

	\begin{center}
		\subfigure[Linear contribution.]{\scalebox{0.25}{\huge \input{Figures/Contributions/ContributionLinearHigh.tex}}} \qquad
		\subfigure[Quadratic contribution.]{\scalebox{0.25}{\huge \input{Figures/Contributions/ContributionQuadraticHigh.tex}}} \\
		\subfigure[Cubic contribution.]{\scalebox{0.25}{\huge \input{Figures/Contributions/ContributionCubicHigh.tex}}} 
		\caption{Confidence bands of the kernels contribution to the total system response obtained through the Volterra model with 99\% of confidence, considering high level of input (1 N). \colorbox{test}{\textcolor{test}{B}} represents the reference condition ($\alpha = 1.00$) and \colorbox{test2}{\textcolor{test2}{B}} the damaged condition ($\alpha = 0.86$).}
		\label{RespChirpContributionHigh}
	\end{center}
\end{figure}

The main advantage in the use of the approach based on Volterra series is the capability to separate linear and nonlinear contributions in the total response, through the Volterra kernels. In this work, the idea is to use this capability to filter and compare the nonlinear contributions to the total system response before and after the damage occurrence. In order to exemplify this information, fig. \ref{RespChirpContributionHigh} shows the contributions of the first, second and third kernels with 99\% of statistical confidence, considering a high level of input signal (1 N) and two structural conditions, healthy ($\alpha = 1.00$) and severe damage ($\alpha = 0.86$). All the contributions, linear, quadratic and cubic, change with the occurrence of the severe damage because the natural frequency and, consequently, the Kautz poles are influenced by the crack behavior. This result shows that the kernels contributions, or kernels coefficients, can be used in the process of damage detection. The variation of the quadratic contribution is bigger, because of the nonlinear nature of the damage. Finally, in the situation of severe damage, the difference between the response in healthy and damaged condition can be visually observed, but at the beginning of the crack propagation, the differentiation is more difficult, mainly in the presence of uncertainties. Therefore, an index has to be used to detect the structural variations. It is expected that the index related with the second kernel will be more sensitive to the presence of the crack.

\FloatBarrier

\subsection{Damage detection}
This section exemplifies the two different metrics proposed to detect damage in nonlinear systems, using the stochastic version of the Volterra series. As mentioned before, the tests were performed considering $\alpha = 1.00,\, 0.98,\, 0.96,\, 0.94$, $\, 0.92,\, 0.90,\, 0.88,\, 0.86$ (2 reference and 7 damage conditions) and the simulated data variation, considering the uncertainties in linear stiffness and damping coefficient. The first 2048 realizations in the reference condition were used as training data and the others as test data. 

\subsubsection{Use of kernels coefficients}
\begin{figure}[!htb]
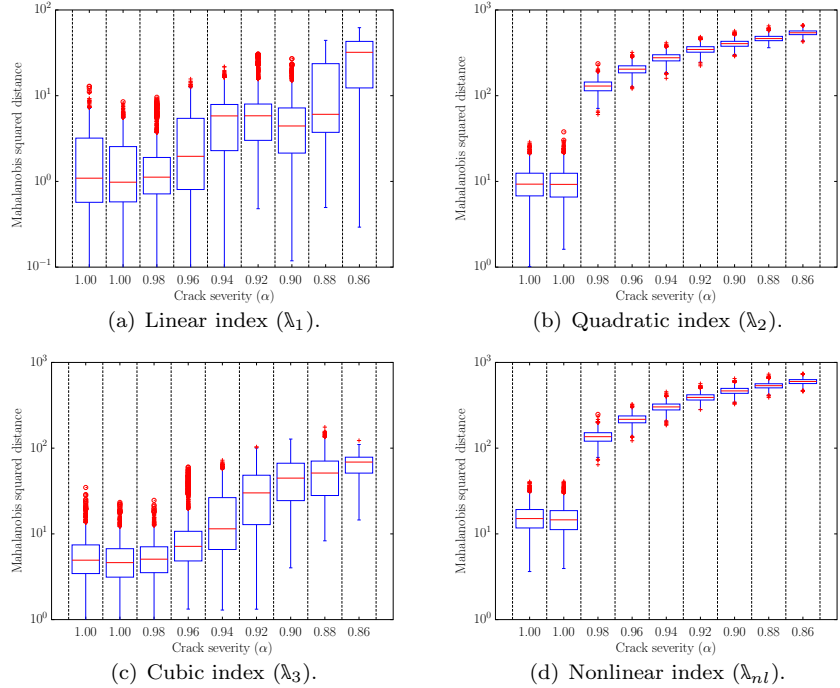

	\begin{center}
		\subfigure[Linear index ($\bblambda_{1}$).]{\scalebox{0.25}{\huge \input{Figures/DamageDetection/IndexLinCoefficients.tex}}} \qquad
		\subfigure[Quadratic index ($\bblambda_{2}$).]{\scalebox{0.25}{\huge \input{Figures/DamageDetection/IndexQuadCoefficients.tex}}} \\
		\subfigure[Cubic index ($\bblambda_{3}$).]{\scalebox{0.25}{\huge \input{Figures/DamageDetection/IndexCubCoefficients.tex}}} \qquad
		\subfigure[Nonlinear index ($\bblambda_{nl}$).]{\scalebox{0.25}{\huge \input{Figures/DamageDetection/IndexNLinCoefficients.tex}}} 
		\caption{Boxplot of the Mahalanobis squared distance applied to the four indexes calculated, for different levels of crack severity.}
		\label{DamageIndexCoefficients}
	\end{center}
\end{figure}

The approach proposed, based on Volterra kernels coefficients, showed in section \ref{Damage detection based on Kernels coefficients} was applied, combined with the novelty detection showed in section \ref{Novelty detection}. The multiple models identified in all structural conditions were considered in the calculation of the indexes. Four different indexes were calculated ($\bblambda_{1}$, $\bblambda_{2}$, $\bblambda_{3}$ and $\bblambda_{nl}$), to compare the results. Figure \ref{DamageIndexCoefficients} shows the evolution of the Mahalanobis squared distance applied to the indexes, with the crack propagation. The boxplot method is used to clarify the visualization of the results. The first two boxes represent the reference condition, the data used to train the model and the data used to test the model. It is observed that the linear index is not able to detect the crack evolution, considering only the kernels coefficients as an indicator. The cubic index has a small increase with the crack evolution, whereas the quadratic index has a larger increase. This occurs because of the quadratic nature of the crack behavior. How expected, the better performance is obtained through the use of the nonlinear index, considering both the coefficients of the second and third kernels.  The dispersion of the values is small, because the uncertainties in the modal parameters have lower influence in the estimation of the higher order Volterra kernels coefficients. The dispersion observed in the quadratic index is related to the noise added because the quadratic component has the level of amplitude close to the noise.

\begin{figure}[!htb]
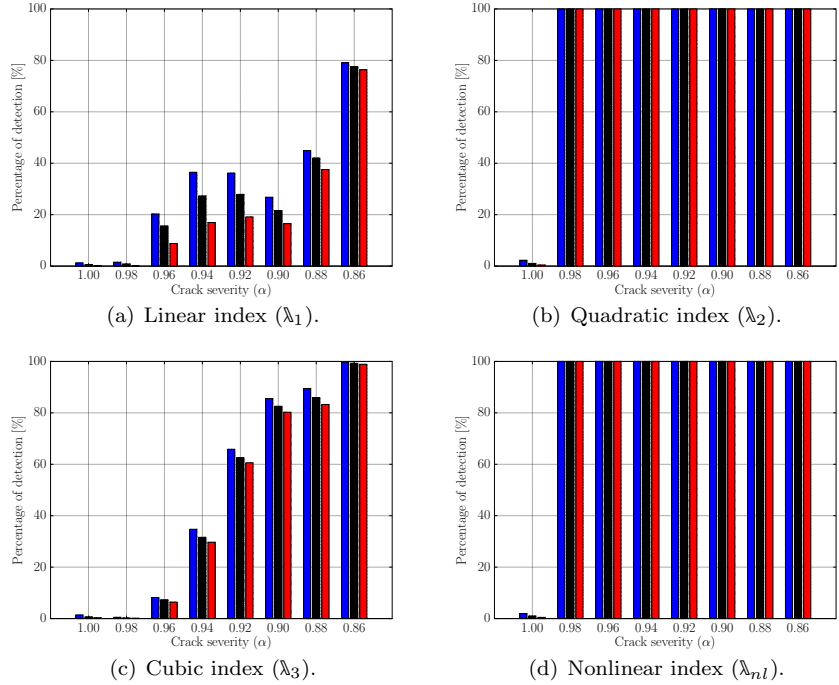

	\begin{center}
		\subfigure[Linear index ($\bblambda_{1}$).]{\scalebox{0.25}{\huge \input{Figures/DamageDetection/PODLinCoefficients.tex}}} \qquad
		\subfigure[Quadratic index ($\bblambda_{2}$).]{\scalebox{0.25}{\huge \input{Figures/DamageDetection/PODQuadCoefficients.tex}}} \\
		\subfigure[Cubic index ($\bblambda_{3}$).]{\scalebox{0.25}{\huge \input{Figures/DamageDetection/PODCubCoefficients.tex}}} \qquad
		\subfigure[Nonlinear index ($\bblambda_{nl}$).]{\scalebox{0.25}{\huge \input{Figures/DamageDetection/PODNLinCoefficients.tex}}} 
		\caption{Percentage of damage detection obtained using the kernels coefficients, considering different crack severities and probability of false alarms used. \colorbox{blue}{\textcolor{blue}{B}} represents $\beta = 0.02$, \colorbox{black}{\textcolor{black}{B}} represents $\beta = 0.01$ and \colorbox{red}{\textcolor{red}{B}} represents $\beta = 0.005$.}
		\label{PODCoefficients}
	\end{center}
\end{figure}

With the Mahalanobis squared distance of the indexes calculated in the reference condition, and considering their empirical distribution obtained, it is possible to define threshold values, taking into account a probability of false alarms ($\beta$) required in each application. In this work, three different values were considered, $\beta = 0.005$, $0.01$ and $0.02$. The results of the hypothesis test applied are shown in fig. \ref{PODCoefficients}, for the three values of $\beta$ used. The performance of the linear index it is not satisfactory, with a low percentage of detection in the initial propagation of the crack. The cubic index can detect the crack depending on the severity of the damage, i. e., for $\alpha$ close to 1.00 the index fails, but for $\alpha \leq 0.86$, all damaged conditions are detected. Finally, the quadratic and nonlinear (sum of quadratic and cubic coefficients) indexes have similar behavior, detecting the damage even on the initial conditions of crack propagation. Remembering that it was considered the data variation, related with the presence of uncertainties, in the process of kernels identification, but, nevertheless, the nonlinear index was able to make difference between the variations related with changes in the natural frequency and damping ratio and the variations related with the crack behavior. This is an important result, showing that the approach can detect damage in uncertain nonlinear systems with probabilistic confidence. 

\begin{figure}[!htb]
	\begin{center}
		\scalebox{0.4}{\huge \input{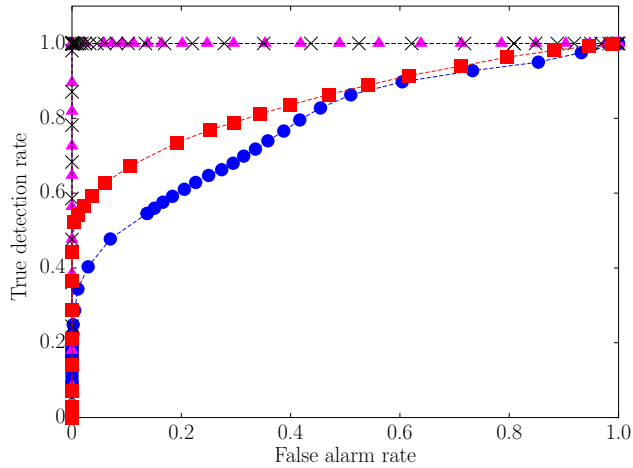}}
		\caption{Receiver operating characteristics (ROC) curve, obtained using the kernels coefficients. \textcolor{blue}{--} \mycircle{blue} \textcolor{blue}{--} represents the linear index, \textcolor{black}{-- x --} the quadratic index, \textcolor{red}{--} \mysquare{red}  \textcolor{red}{--} the cubic index and \textcolor{magenta}{--} \mytriangle{magenta} \textcolor{magenta}{--} the nonlinear index.}
		\label{ROCCoefficients}
	\end{center}
\end{figure}

With the aim of study the performance of the approach proposed, the receiver operating characteristics (ROC) curve was computed \cite{farrar2012structural}. This curve presents the relation between true detections and false alarms for different $\beta$ values. Figure \ref{ROCCoefficients} shows the results obtained for the 4 indexes. The linear index has the worst performance and the cubic index an intermediate performance. The quadratic and nonlinear indexes have similar performance, with a higher rate of detection without expressive false alarms rate. These results show that the nonlinear index proposed is able to detect the initial propagation of the crack even in the presence of uncertainties, simulated by variations in the natural frequency and damping ratio of the system. This high performance is possible because of the nonlinear nature of the damage simulated that has a large influence in the coefficients of the second kernel estimated.  The use of the Volterra series in this situation improve the results in the damage detection process, representing a real contribution when the data variation related to the uncertainties is considered.

\FloatBarrier

\subsubsection{Use of kernels contribution}
Now, the approach proposed based on Volterra kernels contribution showed in section \ref{Damage detection based on Kernels contribution} was applied, combined with the novelty detection showed in section \ref{Novelty detection}. This methodology uses the main advantage of the use of the Volterra series, that is the capability of separate the kernels contribution to the total response. The difference compared with the approach presented in the last section is the influence of the Kautz functions in the process, that introduces the influence of the natural frequency and damping ratio in the approach, through the Kautz poles dependence. The increase of the crack makes the natural frequency of the system change, and this variation can be detected using this approach, but, on the other hand, it is expected that the uncertainties have more influence using this approach, doing the indexes vary more.

\begin{figure}[!htb]
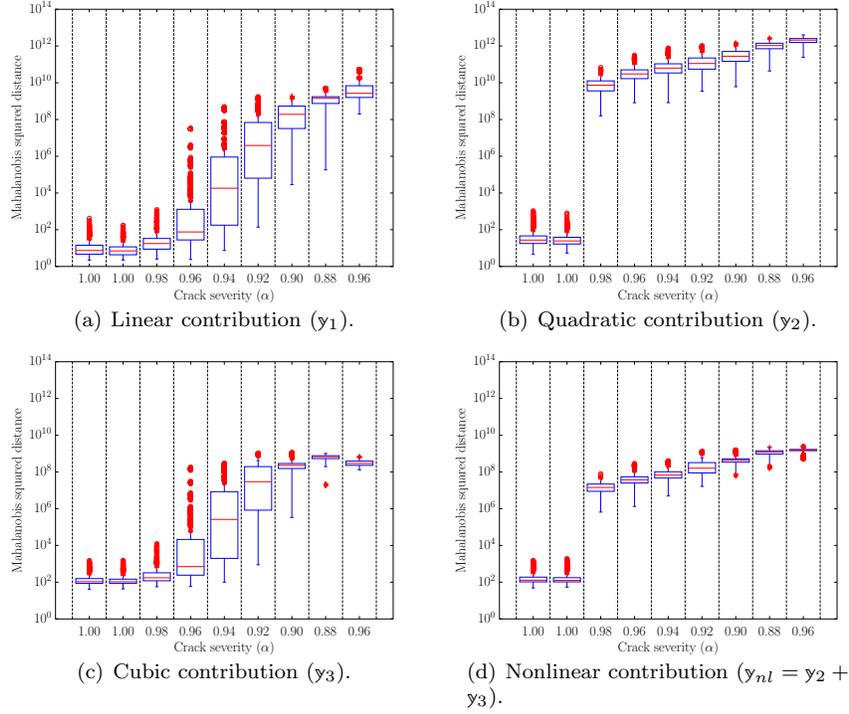

	\begin{center}
		\subfigure[Linear contribution ($\randvar{y}_{1}$).]{\scalebox{0.25}{\huge \input{Figures/DamageDetection/IndexLinContribution.tex}}} \qquad
		\subfigure[Quadratic contribution ($\randvar{y}_{2}$).]{\scalebox{0.25}{\huge \input{Figures/DamageDetection/IndexQuadContribution.tex}}} \\
		\subfigure[Cubic contribution ($\randvar{y}_{3}$).]{\scalebox{0.25}{\huge \input{Figures/DamageDetection/IndexCubContribution.tex}}} \qquad
		\subfigure[Nonlinear contribution ($\randvar{y}_{nl} = \randvar{y}_{2} + \randvar{y}_{3}$).]{\scalebox{0.25}{\huge \input{Figures/DamageDetection/IndexNLinContribution.tex}}} 
		\caption{Boxplot of Mahalanobis squared distance of the kernels contribution calculated for different levels of crack severity.}
		\label{DamageIndexContribution}
	\end{center}
\end{figure}

Again, the multiple models identified in all system conditions (healthy and damaged) were considered in the calculation of the contributions. The input used here is the same chirp signal used in the estimation of the kernels, considering a high level of amplitude (1 N), to maximizes the nonlinear influence in the response. Four different indexes were calculated ($\randvar{y}_{1}$, $\randvar{y}_{2}$, $\randvar{y}_{3}$ and $\randvar{y}_{nl}$), to compare the results.  Figure \ref{DamageIndexContribution} shows the evolution of the Mahalanobis squared distance applied to the kernels contribution, with the crack progression. It is clear that using this approach the linear and cubic indexes have better performance to detect the presence of the crack compared with the use of kernels coefficients. The increase of the indexes is related to the variation of the natural frequency with the progression of the damage. Again, the better results are obtained through the use of the quadratic or nonlinear indexes, because of the nature of the damage simulated. The use of the kernels contribution increases the performance of the linear and cubic indexes, but the performance of the quadratic and nonlinear indexes continued to be better.

With the Mahalanobis squared distance of the kernels contribution calculated in the reference condition, and considering their empirical distribution obtained, it is possible to define threshold values, taking into account the probability of false alarms ($\beta$) required in each application. Again, three different values were considered, $\beta = 0.005$, $0.01$ and $0.02$. The results of the hypothesis test applied are shown in fig. \ref{PODContribution}, for the three values of $\beta$ used. The performance of the linear and cubic indexes was improved with the use of kernels contribution, with a higher percentage of detection of the crack for $\alpha \leq 0.92$. The quadratic and nonlinear (sum of quadratic and cubic contributions) indexes were able to detect the damage even on initial conditions of crack propagation. The variation of the frequency and damping is considered in the Kautz poles estimation, improving the capability of the approach to detect the frequency variation related with the crack behavior, but confusing this variation with the changes that are a consequence of the presence of uncertainties.   

\begin{figure}[!htb]
	\begin{center}
		\subfigure[Linear contribution ($\randvar{y}_{1}$).]{\scalebox{0.25}{\huge \input{Figures/DamageDetection/PODLinContribution.tex}}} \qquad
		\subfigure[Quadratic contribution ($\randvar{y}_{2}$).]{\scalebox{0.25}{\huge \input{Figures/DamageDetection/PODQuadContribution.tex}}} \\
		\subfigure[Cubic contribution ($\randvar{y}_{3}$).]{\scalebox{0.25}{\huge \input{Figures/DamageDetection/PODCubContribution.tex}}} \qquad
		\subfigure[Nonlinear contribution ($\randvar{y}_{nl} = \randvar{y}_{2} + \randvar{y}_{3}$).]{\scalebox{0.25}{\huge \input{Figures/DamageDetection/PODNLinContribution.tex}}} 
		\caption{Percentage of damage detection obtained using the kernels contribution, considering different crack severities and probability of false alarms used. \colorbox{blue}{\textcolor{blue}{B}} represents $\beta = 0.02$, \colorbox{black}{\textcolor{black}{B}} represents $\beta = 0.01$ and \colorbox{red}{\textcolor{red}{B}} represents $\beta = 0.005$.}
		\label{PODContribution}
	\end{center}
\end{figure}

Finally, the receiver operating characteristics (ROC) curve was computed again \cite{farrar2012structural}. Figure \ref{ROCContribution} shows the results obtained in all situations, considering linear and nonlinear kernels. The linear and cubic contributions have the worst performance again because these components of the response have a slight variation for values of crack close to 1.00. The quadratic and nonlinear indexes have close performance, with a higher rate of detection without expressive false alarms rate. These results show that the analysis of the nonlinear contribution to the total response is able to detect the initial propagation of the crack even in the presence of the uncertainties. This higher performance is possible because of the nonlinear nature of the damage simulated, that has a large influence on the nonlinear dynamics of the system. Again, the use of the Volterra series in this situation represents an advantage in the damage detection process, with the improvement in the results in comparison with the linear approach represented by the first kernel contribution.

\begin{figure}[!htb]
	\begin{center}
		\scalebox{0.4}{\huge \input{Figures/DamageDetection/ROCContribution.tex}}
		\caption{Receiver operating characteristics (ROC) curve, obtained using the kernels contribution. \textcolor{blue}{--} \mycircle{blue} \textcolor{blue}{--} represents the linear index, \textcolor{black}{-- x --} the quadratic index, \textcolor{red}{--} \mysquare{red}  \textcolor{red}{--} the cubic index and \textcolor{magenta}{--} \mytriangle{magenta} \textcolor{magenta}{--} the nonlinear index.}
		\label{ROCContribution}
	\end{center}
\end{figure}

\subsubsection{Comparison between the use of kernels coefficients and kernels contribution}
Comparing the results that have been shown in figs. \ref{DamageIndexCoefficients} and \ref{DamageIndexContribution}, it can be noticed that the indexes variation is bigger using the kernels contribution, because the data variation induces the Kautz poles to vary, making the use of kernels contribution more sensitive to the presence of uncertainties than the use of the kernels coefficients. On the other hand, the capability of the Kautz functions to describe the variation of the frequency of oscillation related with the crack presence makes the approach more sensitive to the presence of the damage, mainly considering the linear component of the response. So, damages that induce linear variations in the response, related with frequencies of oscillation or damping ratios, can be better detected using the kernels contribution. Additionally, comparing the figs. \ref{ROCCoefficients} and \ref{ROCContribution} one can observe that the performance of the linear and cubic indexes improves with the use of the kernels contribution, with a higher rate of true detection, but in some sceneries where the presence of the uncertainties can have higher influence in the system response, the confusion between the data variation and the presence of damage must be greater.
	
Now, when the damage induces the system to present nonlinear behavior, like the breathing crack phenomenon studied, the coefficients of the higher order kernels showed to be sensitive to the presence of the damage. In this situation, the approach based on the kernels coefficients is more effective to make difference between the uncertainties and the damages. Therefore, both approaches have shown interesting performance and different characteristics, related to the sensitivity to the presence of the damage and the uncertainties. So, the decision about the use of one or other approach it will depend on the problem considered, given that, when the damage induces nonlinear behavior to the system response, the level of noise it is not so high and the uncertainties are related with linear components, like the example simulated in this paper, the use of the kernels coefficients it is interesting.

\section{Final remarks}
\label{Final Remarks}
The paper has presented an approach to detect damages in initially nonlinear systems, considering data variation caused by the uncertainties. The approach is based on a stochastic version of the Volterra series expanded using random Kautz functions. Two different methodologies were presented, considering the Volterra kernels coefficients and the Volterra kernels contribution. The final decision about the state of the system (healthy or damaged) was performed considering the Mahalanobis squared distance and a threshold value, established from the model estimated in the reference condition. Linear and nonlinear analysis were compared, using the capability of the Volterra series to separate the model's contribution. About the results, the two approaches proposed have presented similar performance. As expected, the linear analysis fails when the damage has low severity, because of the influence of the uncertainties in the system response. The nonlinear analysis uses the nonlinear dynamics of the damage to detect it. It is important to highlight that the nonlinear behavior is considered, in this work, even in the reference condition, which combined with the presence of uncertainties, makes difficult the damage detection process. But, the nonlinear indexes have shown to be able to differ the initial nonlinear behavior and data uncertainties of the crack behavior, with satisfactory performance. As future work, the authors intend to apply the procedure considering an experimental setup composed by a nonlinear beam subject to the presence of a breathing crack.

\section*{Acknowledgments}
The authors are thankful for the financial support provided from of S\~ao Paulo 
Research Foundation (FAPESP), grant number 2012/09135-3 and 2015/25676-2, from Research Support Foundation of the State of 
Rio de Janeiro (FAPERJ), grant number E-26/010.002178/2015, 
and CNPq grant number 307520/2016-1. 

\bibliography{SHMreferences}

\begin{thebibliography}{10}
\expandafter\ifx\csname url\endcsname\relax
  \def\url#1{\texttt{#1}}\fi
\expandafter\ifx\csname urlprefix\endcsname\relax\def\urlprefix{URL }\fi
\expandafter\ifx\csname href\endcsname\relax
  \def\href#1#2{#2} \def\path#1{#1}\fi

\bibitem{Farrar2007}
C.~R. Farrar, K.~Worden, An introduction to structural health monitoring,
  Philosophical Transactions of the Royal Society A: Mathematical,Physical and
  Engineering Sciences 365~(1851) (2007) 303--315.
\newblock \href {http://dx.doi.org/10.1098/rsta.2006.1928}
  {\path{doi:10.1098/rsta.2006.1928}}.

\bibitem{farrar2012structural}
C.~R. Farrar, K.~Worden, Structural health monitoring: a machine learning
  perspective, John Wiley \& Sons, Ltd, 2012.
\newblock \href {http://dx.doi.org/10.1002/9781118443118}
  {\path{doi:10.1002/9781118443118}}.

\bibitem{WORDEN2018139}
K.~Worden, E.~Cross, On switching response surface models, with applications to
  the structural health monitoring of bridges, Mechanical Systems and Signal
  Processing 98 (2018) 139 -- 156.
\newblock \href
  {http://dx.doi.org/http://dx.doi.org/10.1016/j.ymssp.2017.04.022}
  {\path{doi:http://dx.doi.org/10.1016/j.ymssp.2017.04.022}}.

\bibitem{Soize20132379}
C.~Soize, Stochastic modeling of uncertainties in computational structural
  dynamics -- {R}ecent theoretical advances, Journal of Sound and Vibration
  332~(10) (2013) 2379 -- 2395.
\newblock \href {http://dx.doi.org/http://dx.doi.org/10.1016/j.jsv.2011.10.010}
  {\path{doi:http://dx.doi.org/10.1016/j.jsv.2011.10.010}}.

\bibitem{soize2012stochastic}
C.~Soize, Stochastic models of uncertainties in computational mechanics, Amer
  Society of Civil Engineers, 2012.

\bibitem{Soize2017}
C.~Soize, Uncertainty Quantification: An Accelerated Course with Advanced
  Applications in Computational Engineering, Springer, 2017.

\bibitem{MAO2013333}
Z.~Mao, M.~Todd, Statistical modeling of frequency response function estimation
  for uncertainty quantification, Mechanical Systems and Signal Processing
  38~(2) (2013) 333 -- 345.
\newblock \href {http://dx.doi.org/https://doi.org/10.1016/j.ymssp.2013.01.021}
  {\path{doi:https://doi.org/10.1016/j.ymssp.2013.01.021}}.

\bibitem{WORDEN20151}
K.~Worden, E.~J. Cross, N.~Dervilis, E.~Papatheou, I.~Antoniadou, Structural
  {H}ealth {M}onitoring: from {S}tructures to {S}ystems-of-{S}ystems.,
  IFAC-PapersOnLine 48~(21) (2015) 1 -- 17, 9th IFAC Symposium on Fault
  Detection, Supervision andSafety for Technical Processes SAFEPROCESS 2015.
\newblock \href
  {http://dx.doi.org/http://dx.doi.org/10.1016/j.ifacol.2015.09.497}
  {\path{doi:http://dx.doi.org/10.1016/j.ifacol.2015.09.497}}.

\bibitem{Pavlopoulou2016}
S.~Pavlopoulou, K.~Worden, C.~Soutis, Novelty detection and dimension reduction
  via guided ultrasonic waves: {D}amage monitoring of scarf repairs in
  composite laminates, Journal of Intelligent Material Systems and Structures
  27~(4) (2016) 549--566.
\newblock \href {http://dx.doi.org/10.1177/1045389X15574937}
  {\path{doi:10.1177/1045389X15574937}}.

\bibitem{sohn2005structural}
H.~Sohn, D.~W. Allen, K.~Worden, C.~R. Farrar, Structural damage classification
  using extreme value statistics, Journal of dynamic systems, measurement, and
  control 127~(1) (2005) 125 -- 132.
\newblock \href {http://dx.doi.org/10.1115/1.1849240}
  {\path{doi:10.1115/1.1849240}}.

\bibitem{MarcRebillat2016}
M.~R\'ebillat, O.~Hmad, F.~Kadri, N.~Mechbal, Peaks {O}ver {T}hreshold-based
  detector design for structural health monitoring: {A}pplication to aerospace
  structures, Structural Health Monitoring 0~(0) (2017) 1475--9217.
\newblock \href {http://dx.doi.org/10.1177/1475921716685039}
  {\path{doi:10.1177/1475921716685039}}.

\bibitem{SANTOS2016584}
A.~Santos, E.~Figueiredo, M.~Silva, C.~Sales, J.~Costa, Machine learning
  algorithms for damage detection: {K}ernel-based approaches, Journal of Sound
  and Vibration 363 (2016) 584 -- 599.
\newblock \href {http://dx.doi.org/http://dx.doi.org/10.1016/j.jsv.2015.11.008}
  {\path{doi:http://dx.doi.org/10.1016/j.jsv.2015.11.008}}.

\bibitem{WORDEN2003323}
K.~WORDEN, G.~MANSON, D.~ALLMAN, Experimental validation of a structural health
  monitoring methodology: Part i. novelty detection on a laboratory structure,
  Journal of Sound and Vibration 259~(2) (2003) 323 -- 343.
\newblock \href {http://dx.doi.org/http://dx.doi.org/10.1006/jsvi.2002.5168}
  {\path{doi:http://dx.doi.org/10.1006/jsvi.2002.5168}}.

\bibitem{FIGUEIREDO20141}
E.~Figueiredo, L.~Radu, K.~Worden, C.~R. Farrar, A bayesian approach based on a
  {M}arkov-{C}hain {M}onte {C}arlo method for damage detection under unknown
  sources of variability, Engineering Structures 80 (2014) 1 -- 10.
\newblock \href
  {http://dx.doi.org/http://dx.doi.org/10.1016/j.engstruct.2014.08.042}
  {\path{doi:http://dx.doi.org/10.1016/j.engstruct.2014.08.042}}.

\bibitem{WORDEN2000647}
K.~WORDEN, G.~MANSON, N.~FIELLER, Damage detection using outlier analysis,
  Journal of Sound and Vibration 229~(3) (2000) 647 -- 667.
\newblock \href {http://dx.doi.org/http://dx.doi.org/10.1006/jsvi.1999.2514}
  {\path{doi:http://dx.doi.org/10.1006/jsvi.1999.2514}}.

\bibitem{FIGUEIREDO2010822}
E.~Figueiredo, M.~Todd, C.~Farrar, E.~Flynn, Autoregressive modeling with
  state-space embedding vectors for damage detection under operational
  variability, International Journal of Engineering Science 48~(10) (2010) 822
  -- 834, structural Health Monitoring in the Light of Inverse Problems of
  Mechanics.
\newblock \href
  {http://dx.doi.org/https://doi.org/10.1016/j.ijengsci.2010.05.005}
  {\path{doi:https://doi.org/10.1016/j.ijengsci.2010.05.005}}.

\bibitem{LIM2014468}
H.~J. Lim, H.~Sohn, M.~P. DeSimio, K.~Brown, Reference-free fatigue crack
  detection using nonlinear ultrasonic modulation under various temperature and
  loading conditions, Mechanical Systems and Signal Processing 45~(2) (2014)
  468 -- 478.
\newblock \href
  {http://dx.doi.org/http://dx.doi.org/10.1016/j.ymssp.2013.12.001}
  {\path{doi:http://dx.doi.org/10.1016/j.ymssp.2013.12.001}}.

\bibitem{Mandal2015}
D.~D. Mandal, D.~Wadadar, S.~Banerjee, Health monitoring of stiffened metallic
  plates using nonlinear wave interaction and embedded pzt transducers, in:
  J.~K. Sinha (Ed.), Vibration Engineering and Technology of Machinery,
  Springer International Publishing, Cham, 2015, pp. 629--638.

\bibitem{ZENG2015380}
M.~Zeng, Y.~Yang, J.~Zheng, J.~Cheng, Normalized complex {T}eager energy
  operator demodulation method and its application to fault diagnosis in a
  rubbing rotor system, Mechanical Systems and Signal Processing 50 (2015) 380
  -- 399.
\newblock \href
  {http://dx.doi.org/http://dx.doi.org/10.1016/j.ymssp.2014.05.006}
  {\path{doi:http://dx.doi.org/10.1016/j.ymssp.2014.05.006}}.

\bibitem{STC:STC215}
K.~Worden, C.~R. Farrar, J.~Haywood, M.~Todd, A review of nonlinear dynamics
  applications to structural health monitoring, Structural Control and Health
  Monitoring 15~(4) (2008) 540--567.
\newblock \href {http://dx.doi.org/10.1002/stc.215}
  {\path{doi:10.1002/stc.215}}.

\bibitem{ANDREAUS2012382}
U.~Andreaus, P.~Baragatti, Experimental damage detection of cracked beams by
  using nonlinear characteristics of forced response, Mechanical Systems and
  Signal Processing 31 (2012) 382 -- 404.
\newblock \href
  {http://dx.doi.org/http://dx.doi.org/10.1016/j.ymssp.2012.04.007}
  {\path{doi:http://dx.doi.org/10.1016/j.ymssp.2012.04.007}}.

\bibitem{REBILLAT2014247}
M.~R\'ebillat, R.~Hajrya, N.~Mechbal, Nonlinear structural damage detection
  based on cascade of {H}ammerstein models, Mechanical Systems and Signal
  Processing 48~(1) (2014) 247 -- 259.
\newblock \href
  {http://dx.doi.org/http://dx.doi.org/10.1016/j.ymssp.2014.03.009}
  {\path{doi:http://dx.doi.org/10.1016/j.ymssp.2014.03.009}}.

\bibitem{Bornn2010909}
L.~Bornn, C.~R. Farrar, G.~Park, Damage detection in initially nonlinear
  systems, International Journal of Engineering Science 48~(10) (2010) 909 --
  920, structural Health Monitoring in the Light of Inverse Problems of
  Mechanics.
\newblock \href {http://dx.doi.org/10.1016/j.ijengsci.2010.05.011}
  {\path{doi:10.1016/j.ijengsci.2010.05.011}}.

\bibitem{Virgin}
L.~N. Virgin, Introduction to Experimental Nonlinear Dynamics, Cambridge
  University Press, 2000.

\bibitem{noel2017nonlinear}
J.-P. No{\"e}l, G.~Kerschen, Nonlinear system identification in structural
  dynamics: 10 more years of progress, Mechanical Systems and Signal Processing
  83 (2016) 2 -- 35.
\newblock \href {http://dx.doi.org/10.1016/j.ymssp.2016.07.020i}
  {\path{doi:10.1016/j.ymssp.2016.07.020i}}.

\bibitem{kerschen2007nonlinear}
G.~Kerschen, K.~Worden, A.~F. Vakakis, J.-C. Golinval, Nonlinear system
  identification in structural dynamics: current status and future directions,
  in: 25th International Modal Analysis Conference, Orlando, 2007, 2007, pp.
  2413--2438.

\bibitem{worden2007nonlinear}
K.~Worden, G.~Kerschen, A.~F. Vakakis, J.-C. Golinval, Nonlinear system
  identification in structural dynamics: A short (and biased) history, in: 25th
  International Modal Analysis Conference, Orlando, 2007, 2007, pp. 1996--2017.

\bibitem{masri1979nonparametric}
S.~Masri, T.~Caughey, A nonparametric identification technique for nonlinear
  dynamic problems, Journal of Applied Mechanics 46~(2) (1979) 433--447.
\newblock \href {http://dx.doi.org/10.1115/1.3424568}
  {\path{doi:10.1115/1.3424568}}.

\bibitem{Noel2012}
J.~P. Noel, G.~Kerschen, A.~Newerla, Application of the restoring force surface
  method to a real-life spacecraft structure, in: D.~Adams, G.~Kerschen,
  A.~Carrella (Eds.), Topics in Nonlinear Dynamics, Volume 3, Springer New
  York, New York, NY, 2012, pp. 1--19.

\bibitem{Tang16092015}
B.~Tang, M.~Brennan, V.~Lopes, S.~da~Silva, R.~Ramlan, Using nonlinear jumps to
  estimate cubic stiffness nonlinearity: {A}n experimental study, Proceedings
  of the Institution of Mechanical Engineers, Part C: Journal of Mechanical
  Engineering Science\href {http://dx.doi.org/10.1177/0954406215606746}
  {\path{doi:10.1177/0954406215606746}}.

\bibitem{Schetzen}
M.~Schetzen, The {V}olterra and {W}iener Theories of Nonlinear Systems, Krieger
  Publishing Co., Inc., Melbourne, FL, USA, 2006.

\bibitem{CHATTERJEE20101716}
A.~Chatterjee, Parameter estimation of {D}uffing oscillator using {V}olterra
  series and multi-tone excitation, International Journal of Mechanical
  Sciences 52~(12) (2010) 1716 -- 1722.
\newblock \href
  {http://dx.doi.org/http://dx.doi.org/10.1016/j.ijmecsci.2010.09.005}
  {\path{doi:http://dx.doi.org/10.1016/j.ijmecsci.2010.09.005}}.

\bibitem{Scussel2017}
O.~Scussel, S.~da~Silva, The harmonic probing method for output-only nonlinear
  mechanical systems, Journal of the Brazilian Society of Mechanical Sciences
  and Engineering\href {http://dx.doi.org/10.1007/s40430-017-0723-y}
  {\path{doi:10.1007/s40430-017-0723-y}}.

\bibitem{Chatterjee20103325}
A.~Chatterjee, Structural damage assessment in a cantilever beam with a
  breathing crack using higher order frequency response functions, Journal of
  Sound and Vibration 329~(16) (2010) 3325 -- 3334.
\newblock \href {http://dx.doi.org/http://dx.doi.org/10.1016/j.jsv.2010.02.026}
  {\path{doi:http://dx.doi.org/10.1016/j.jsv.2010.02.026}}.

\bibitem{TANG20101099}
H.~Tang, Y.~Liao, J.~Cao, H.~Xie, Fault diagnosis approach based on volterra
  models, Mechanical Systems and Signal Processing 24~(4) (2010) 1099 -- 1113.
\newblock \href {http://dx.doi.org/https://doi.org/10.1016/j.ymssp.2009.09.001}
  {\path{doi:https://doi.org/10.1016/j.ymssp.2009.09.001}}.

\bibitem{XIA2016557}
X.~Xia, J.~Zhou, J.~Xiao, H.~Xiao, A novel identification method of {V}olterra
  series in rotor-bearing system for fault diagnosis, Mechanical Systems and
  Signal Processing 66-67 (2016) 557 -- 567.
\newblock \href {http://dx.doi.org/https://doi.org/10.1016/j.ymssp.2015.05.006}
  {\path{doi:https://doi.org/10.1016/j.ymssp.2015.05.006}}.

\bibitem{BruceSHMjournal}
S.~B. Shiki, S.~da~Silva, M.~D. Todd, On the application of discrete-time
  {V}olterra series for the damage detection problem in initially nonlinear
  systems, Structural Health Monitoring 16~(1) (2017) 62--78.
\newblock \href {http://dx.doi.org/10.1177/1475921716662142}
  {\path{doi:10.1177/1475921716662142}}.

\bibitem{Kautz}
W.~H. Kautz, Transient synthesis in the time domain, Circuit Theory,
  Transactions of the IRE Professional Group on CT-1~(3) (1954) 29--39.
\newblock \href {http://dx.doi.org/10.1109/TCT.1954.1083588}
  {\path{doi:10.1109/TCT.1954.1083588}}.

\bibitem{heuberger2005modelling}
P.~S. Heuberger, P.~M. Van~den Hof, B.~Wahlberg, Modelling and identification
  with rational orthogonal basis functions, Springer, 2005.

\bibitem{daSilva20111103}
S.~da~Silva, Non-parametric identification of mechanical systems by {K}autz
  filter with multiple poles, Mechanical Systems and Signal Processing 25~(4)
  (2011) 1103 -- 1111.
\newblock \href {http://dx.doi.org/10.1016/j.ymssp.2010.11.010}
  {\path{doi:10.1016/j.ymssp.2010.11.010}}.

\bibitem{daSilva2011312}
S.~da~Silva, Non-linear model updating of a three-dimensional portal frame
  based on {W}iener series, International Journal of Non-Linear Mechanics
  46~(1) (2011) 312 -- 320.
\newblock \href {http://dx.doi.org/10.1016/j.ijnonlinmec.2010.09.014}
  {\path{doi:10.1016/j.ijnonlinmec.2010.09.014}}.

\bibitem{WAHLBERG1996693}
B.~Wahlberg, P.~Makila, On approximation of stable linear dynamical systems
  using {L}aguerre and {K}autz functions, Automatica 32~(5) (1996) 693 -- 708.
\newblock \href
  {http://dx.doi.org/http://dx.doi.org/10.1016/0005-1098(95)00198-0}
  {\path{doi:http://dx.doi.org/10.1016/0005-1098(95)00198-0}}.

\bibitem{Oliveira2012}
G.~H. Oliveira, A.~da~Rosa, R.~J. Campello, J.~B. Machado, W.~C. Amaral, An
  introduction to models based on {L}aguerre, {K}autz and other related
  orthonormal functions - part {I}{I}: non-linear models, International Journal
  of Modelling, Identification and Control 16~(1) (2012) 1--14.
\newblock \href {http://dx.doi.org/10.1504/IJMIC.2012.046691}
  {\path{doi:10.1504/IJMIC.2012.046691}}.

\bibitem{Wahlberg1994}
B.~Wahlberg, System identification using {K}autz models, IEEE Transactions on
  Automatic Control 39~(6) (1994) 1276--1282.
\newblock \href {http://dx.doi.org/10.1109/9.293196}
  {\path{doi:10.1109/9.293196}}.

\bibitem{AdaRosa}
A.~da~Rosa, R.~J. Campello, W.~C. Amaral, Choice of free parameters in
  expansions of discrete-time {V}olterra models using {K}autz functions,
  Automatica 43~(6) (2007) 1084 -- 1091.
\newblock \href
  {http://dx.doi.org/http://dx.doi.org/10.1016/j.automatica.2006.12.007}
  {\path{doi:http://dx.doi.org/10.1016/j.automatica.2006.12.007}}.

\bibitem{silverman1986density}
B.~W. Silverman, Density estimation for statistics and data analysis, Vol.~26,
  CRC press, 1986.

\bibitem{Bowman1997}
A.~W. Bowman, A.~Azzalini, Applied smoothing techniques for data analysis: the
  kernel approach with S-Plus illustrations, Vol.~18, OUP Oxford, 1997.

\bibitem{kovacic2011duffing}
I.~Kovacic, M.~J. Brennan, The Duffing equation: nonlinear oscillators and
  their behaviour, John Wiley \& Sons, 2011.

\bibitem{CHATI1997249}
M.~Chati, R.~Rand, S.~Mukherjee, Modal analysis of a cracked beam, Journal of
  Sound and Vibration 207~(2) (1997) 249 -- 270.
\newblock \href {http://dx.doi.org/http://dx.doi.org/10.1006/jsvi.1997.1099}
  {\path{doi:http://dx.doi.org/10.1006/jsvi.1997.1099}}.

\bibitem{ANDREAUS2007566}
U.~Andreaus, P.~Casini, F.~Vestroni, Non-linear dynamics of a cracked
  cantilever beam under harmonic excitation, International Journal of
  Non-Linear Mechanics 42~(3) (2007) 566 -- 575.
\newblock \href
  {http://dx.doi.org/http://dx.doi.org/10.1016/j.ijnonlinmec.2006.08.007}
  {\path{doi:http://dx.doi.org/10.1016/j.ijnonlinmec.2006.08.007}}.

\bibitem{Prawin2018}
J.~Prawin, A.~R.~M. Rao, Nonlinear structural damage detection based on
  adaptive {V}olterra filter model, International Journal of Structural
  Stability and Dynamics 18~(0) (2018) 12.
\newblock \href {http://dx.doi.org/10.1142/S0219455418710037}
  {\path{doi:10.1142/S0219455418710037}}.

\bibitem{PENG2007777}
Z.~Peng, Z.~Lang, S.~Billings, Crack detection using nonlinear output frequency
  response functions, Journal of Sound and Vibration 301~(3) (2007) 777 -- 788.
\newblock \href {http://dx.doi.org/https://doi.org/10.1016/j.jsv.2006.10.039}
  {\path{doi:https://doi.org/10.1016/j.jsv.2006.10.039}}.

\bibitem{Surace2011}
C.~Surace, R.~Ruotolo, D.~Storer, Detecting nonlinear behaviour using the
  {V}olterra series to assess damage in beam-like structures, Journal of
  Theoretical and Applied Mechanics Vol. 49 nr 3 (2011) 905--926.

\bibitem{CunhaJr2017252}
A.~{Cunha~Jr}, J.~L.~P. Felix, J.~M. Balthazar, Quantification of parametric
  uncertainties induced by irregular soil loading in orchard tower sprayer
  nonlinear dynamics, Journal of Sound and Vibration 408 (2017) 252 -- 269.
\newblock \href {http://dx.doi.org/https://doi.org/10.1016/j.jsv.2017.07.023}
  {\path{doi:https://doi.org/10.1016/j.jsv.2017.07.023}}.

\bibitem{Sohn539}
H.~Sohn, Effects of environmental and operational variability on structural
  health monitoring, Philosophical Transactions of the Royal Society of London
  A: Mathematical, Physical and Engineering Sciences 365~(1851) (2007)
  539--560.
\newblock \href {http://dx.doi.org/10.1098/rsta.2006.1935}
  {\path{doi:10.1098/rsta.2006.1935}}.

\bibitem{PhysRev}
E.~T. Jaynes, Information theory and statistical mechanics, Phys. Rev. 106
  (1957) 620--630.
\newblock \href {http://dx.doi.org/10.1103/PhysRev.106.620}
  {\path{doi:10.1103/PhysRev.106.620}}.

\bibitem{cunhajr2017}
A.~{Cunha~Jr}, {M}odeling and {Q}uantification of {P}hysical {S}ystems
  {U}ncertainties in a {P}robabilistic {F}ramework, in: S.~Ekwaro-Osire, A.~C.
  Goncalves, F.~M. Alemayehu (Eds.), Probabilistic Prognostics and Health
  Management of Energy Systems, Springer International Publishing, New York,
  2017, pp. 127--156.

\bibitem{CunhaJr2015809}
A.~Cunha~Jr, R.~Sampaio, On the nonlinear stochastic dynamics of a continuous
  system with discrete attached elements, Applied Mathematical Modelling 39~(2)
  (2015) 809 -- 819.
\newblock \href {http://dx.doi.org/http://dx.doi.org/10.1016/j.apm.2014.07.012}
  {\path{doi:http://dx.doi.org/10.1016/j.apm.2014.07.012}}.

\bibitem{rubinstein2016simulation}
R.~Y. Rubinstein, D.~P. Kroese, Simulation and the {M}onte {C}arlo {M}ethod,
  3rd Edition, Wiley, 2016.

\bibitem{Soize2005623}
C.~Soize, A comprehensive overview of a non-parametric probabilistic approach
  of model uncertainties for predictive models in structural dynamics, Journal
  of Sound and Vibration 288~(3) (2005) 623 -- 652, uncertainty in structural
  dynamics.
\newblock \href {http://dx.doi.org/http://dx.doi.org/10.1016/j.jsv.2005.07.009}
  {\path{doi:http://dx.doi.org/10.1016/j.jsv.2005.07.009}}.

\end{thebibliography}

\end{document}